\documentclass[aps,pre,twocolumn,groupedaddress,longbibliography,10pt,nobibnotes,nofootinbib,floatfix]{revtex4-2}  
\pdfoutput=1

\usepackage{etoolbox}
\usepackage{geometry}
\usepackage{graphicx}%
\usepackage{xr-hyper}%
\usepackage{amssymb}
\usepackage{amsmath}
\usepackage{mathtools}
\usepackage{MnSymbol}
\usepackage[rgb]{xcolor}
\usepackage{float}
\usepackage[title,titletoc]{appendix}
\usepackage{adjustbox}
\newcommand{\aligntop}[1]{\adjustbox{valign=t}{#1}}
\usepackage{hyperref}%
\usepackage{cleveref}

\newcommand{\sumB}{\sum_{i\in\CN,j\in\CS}}

\newcommand{\inv}{^{-1}}

\newcommand{\trans}{^\mathrm{T}}
\newcommand{\dg}{^\mathrm{dg}}
\newcommand{\I}{\mathbf{I}}
\newcommand{\onev}{\mathbf{e}}

\newcommand{\MMm}{Q}
\newcommand{\MM}{\mathbf{\MMm}}
\newcommand{\Mdm}{M}
\newcommand{\Md}{\mathbf{\Mdm}}
\newcommand{\eq}{\boldsymbol{\pi}}
\newcommand{\Eq}{\boldsymbol{\Pi}}
\newcommand{\pr}{\mathbf{p}}
\newcommand{\jb}{\mathbf{j}}
\newcommand{\jbp}{\ssv{\jb}}

\newcommand{\fpt}{\mathbf{T}}
\newcommand{\fptb}{\overline{\fpt}}
\newcommand{\fptbm}{\overline{T}}

\newcommand{\Pbm}{P}
\newcommand{\Pb}{\mathbf{\Pbm}}

\newcommand{\Lb}{\boldsymbol{\Lambda}}
\newcommand{\dinv}{^{\mathcal{D}}}
\newcommand{\dinvd}{^{\mathcal{D}\dag}}

\newcommand{\uv}{\mathbf{u}}
\newcommand{\vv}{\mathbf{v}}
\newcommand{\rowv}{\boldsymbol{\xi}}

\newcommand{\bgm}{\boldsymbol{\gamma}}

\newcommand{\M}{\mathbf{M}}

\newcommand{\pdiff}[3][]{\frac{\partial^{#1} #2}{\partial {#3}^{#1}}}
\newcommand{\ie}{i.e.\ }

\newcommand{\Prob}{\mathbb{P}}
\newcommand{\event}[1]{\left( #1 \right)}
\newcommand{\events}[1]{\left[ #1 \right]}
\newcommand{\cond}[2]{\!\left( #1 \middle\mvert\mathopen{} #2 \right)}
\newcommand{\conds}[2]{\!\left[ #1 \middle\mvert\mathopen{} #2 \right]}

\newcommand{\traj}{x(t)}
\newcommand{\trajs}{\{x, \tau \}}
\newcommand{\otraj}{\event{\trajs}}

\newcommand{\kt}{k_{\mathrm{B}}T}
\newcommand{\CN}{\mathcal{N}}
\newcommand{\CS}{\mathcal{S}}
\newcommand{\CB}{\mathcal{B}}
\newcommand{\CNj}{\vec{\CN}}
\newcommand{\CSj}{\vec{\CS}}
\newcommand{\CBj}{\vec{\CB}}

\newcommand{\lbj}{_{\CBj}}

\newcommand{\Bind}{R}
\newcommand{\ssv}[2][\pi]{#2^{#1}}
\newcommand{\Nb}{\overline{N}}
\newcommand{\qp}{\ssv{q}}
\newcommand{\qm}{1-\qp}
\newcommand{\qmb}{(\qm)}
\newcommand{\qpm}{\qp\qmb}
\newcommand{\qpmb}{\left[\qp\qmb\right]}
\newcommand{\phip}{\ssv{\phi}}

\newcommand{\Sigmap}{\ssv{\Sigma}}
\newcommand{\sigmap}{\ssv{\sigma}}
\newcommand{\jp}{\ssv{j}}

\newcommand{\Bindp}{\ssv{\Bind}}
\newcommand{\emv}[1]{#1^T}
\newcommand{\qem}{\emv{q}}
\newcommand{\pem}{\emv{p}}
\newcommand{\phiem}{\emv{\phi}}
\newcommand{\jem}{\emv{j}}
\newcommand{\fem}{\emv{f}}
\newcommand{\Bindem}{\emv{\Bind}}
\newcommand{\emp}[2][p]{\ssv[#1]{#2}}
\newcommand{\phipem}{\emp{\phi}}
\newcommand{\Bindpem}{\emp{\Bind}}
\newcommand{\jpem}{\emp{j}}
\newcommand{\fpem}{\emp{f}}

\newcommand{\CO}{\mathcal{O}}
\newcommand{\CL}{\mathcal{L}}
\DeclareMathOperator{\Tr}{Tr}
\DeclareMathOperator{\diag}{diag}
\DeclareMathOperator{\arcsinh}{arcsinh}

\newcommand{\e}{\mathrm{e}}

\newcommand{\dt}{\mathrm{d}t}
\newcommand{\dtau}{\mathrm{d}\tau}
\newcommand{\dx}{\mathrm{d}x}
\newcommand{\dtraj}{\{\dx, \dtau \}}
\newcommand{\diff}[3][]{\frac{\mathrm{d}^{#1} #2}{\mathrm{d}{#3}^{#1}}}
\newcommand{\Var}[1]{\av{(\delta #1)^2}}
\newcommand{\tunbind}{T_\text{unbind}}

\newcommand{\thold}{T_\text{hold}}

\newcommand{\object}{receptor} 
\newcommand{\objects}{receptors} 
\newcommand{\Object}{Receptor} 
\newcommand{\Objects}{Receptors} 

\newcommand{\sigfrac}{signaling density}

\newcommand{\error}{fractional error}

\newcommand{\energy}{energy consumption}

\newcommand{\obstraj}{ideal observer}

\newcommand{\IO}{IO}

\newcommand{\obsq}{simple observer}

\newcommand{\SO}{SO}

\newcommand{\prn}[1]{\left ( #1 \right )}
\newcommand{\brc}[1]{\left\{ #1 \right\}}  
\newcommand{\brk}[1]{\left [ #1 \right ]}
\newcommand{\av}[1]{\left\langle #1 \right\rangle}
\newcommand{\abs}[1]{\left\lvert #1 \right\rvert}
\newcommand{\nrm}[1]{\left\lVert #1 \right\rVert}

\AtBeginEnvironment{appendices}{%
  \crefalias{section}{appendix}
  \crefalias{subsection}{subappendix}
  \crefalias{subsubsection}{subsubappendix}
  \phantomsection}
\crefformat{section}{\S#2#1#3}
\crefformat{appendix}{\S#2#1#3}
\crefrangeformat{section}{\S\S#3#1#4--#5\crefstripprefix{#1}{#2}#6}
\crefrangeformat{appendix}{\S\S#3#1#4--#5\crefstripprefix{#1}{#2}#6}

\makeatletter
\renewcommand{\p@subsection}{\thesection.}
\makeatother
\AtBeginEnvironment{appendices}{%
  
}

\begin{document}

\newgeometry{margin=0.75in}

\title{Universal energy-accuracy tradeoffs in nonequilibrium cellular sensing}

\author{Sarah E.~Harvey}
\email[]{harveys@stanford.edu}
\author{Subhaneil~Lahiri}
\author{Surya~Ganguli}
\affiliation{Department of Applied Physics, Stanford University, Stanford, California, 94305}


\date{\today}

\begin{abstract}
We combine stochastic thermodynamics, large deviation theory, and information theory to derive fundamental limits on the accuracy with which single cell receptors can estimate external concentrations.  
  As expected, if estimation is performed by an ideal observer of the entire trajectory of receptor states, then no energy consuming non-equilibrium receptor that can be divided into bound and unbound states can outperform an equilibrium two-state receptor.
  However, when estimation is performed by a simple observer that measures the fraction of time the receptor is bound, we derive a fundamental limit on the accuracy of general non-equilibrium receptors as a function of energy consumption.
We further derive and exploit explicit formulas to numerically estimate a Pareto-optimal tradeoff between accuracy and energy. 
We find this tradeoff can be achieved by nonuniform ring receptors with a number of states that necessarily increases with energy. 
Our results yield a novel thermodynamic uncertainty relation for the time a physical system spends in a pool of states, and generalize the classic 1977 Berg-Purcell limit on cellular sensing along multiple dimensions.
\end{abstract}


\maketitle



\begin{figure}[hb]
  \centering
  \includegraphics[width=1\linewidth]{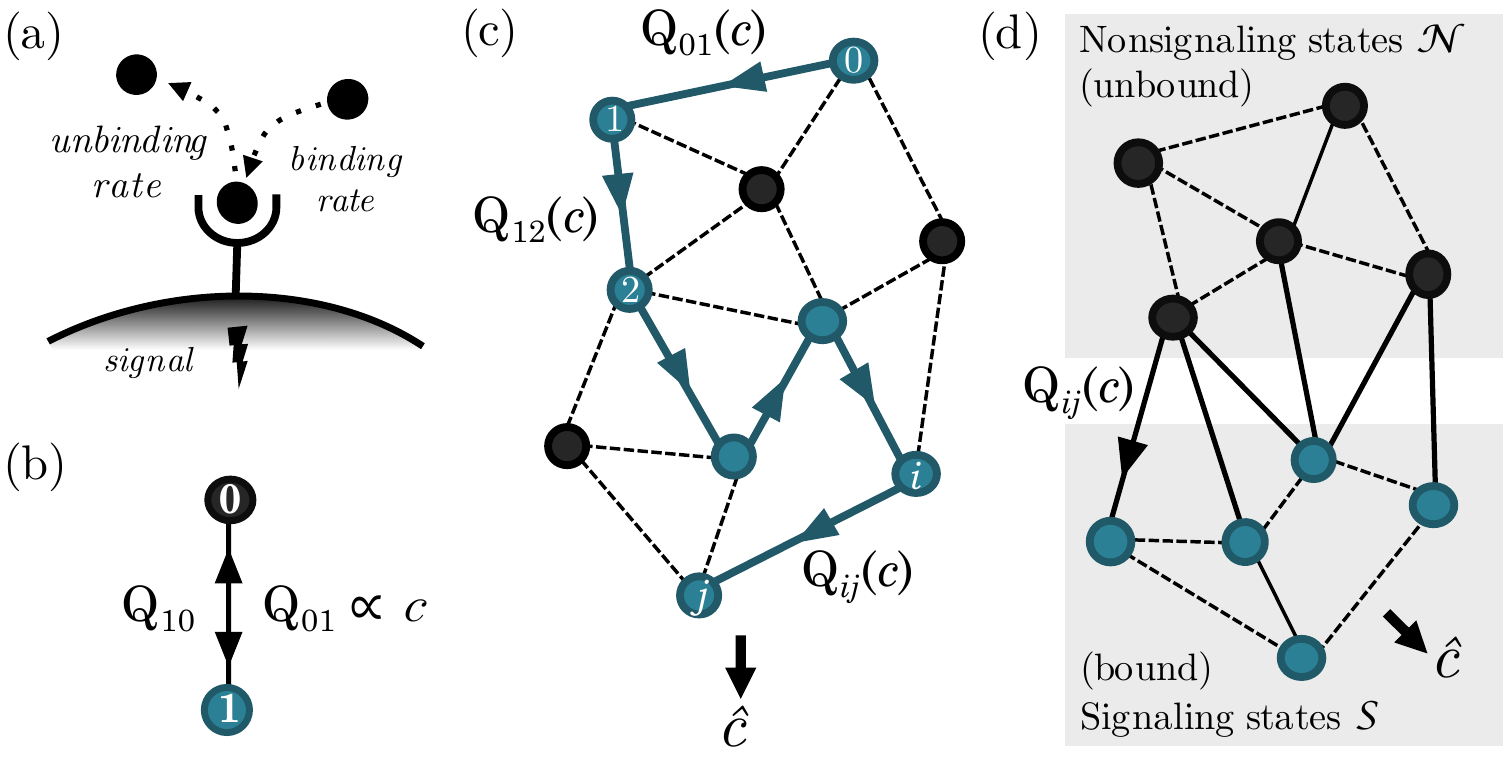}
  \caption[Cartoon of a single-cell chemosensitivity system]
  {Single receptors as continuous-time Markov processes.
    (a)  Cartoon of a single receptor.
    (b) Single receptor modeled as a two-state Markov process.
    (c-d) Generalizing to many-state processes. (c) An \obstraj{} uses the process trajectory to form an estimate \(\hat{c}\) of the concentration signal \(c\), which modulates the receptor's transition rates.
    (d) A \obsq{} uses the fraction of time the receptor spends in a subset of states (`signaling states') to estimate the signal. }\label{fig:fig1}
\end{figure}

Single cells possess extremely sensitive mechanisms for detecting chemical concentrations through the binding of molecules to cell-surface receptors (\cref{fig:fig1}(a),~\cite{Smith2008, Sourjik2012, Bialek2012}). 
This remarkable capacity may require energy consumption, and raises important questions about fundamental limits on the accuracy of cellular chemosensation, both as a function of the energy consumed by arbitrarily complex nonequilibrium  receptors, and the computational sophistication of downstream observers of these dynamics.

A seminal line of work by Berg and Purcell (1977)~\cite{Berg1977, Bialek2005,Kaizu2014} addressed this question for equilibrium receptors with two states, bound and unbound, with the binding transition rate proportional to the external concentration $c$ (\cref{fig:fig1}(b)).
They studied the accuracy of a concentration estimate  $\hat{c}$ computed by a \obsq{} (\SO) that only has access to the fraction of time the receptor is bound over a time $T$, finding a fundamental lower bound on the fractional error of this estimate:
\begin{equation}\label{eq:BergPurcell}
  \epsilon_{\hat{c}}^{2}
  \equiv \frac{\Var{\hat{c}}}{c^{2}}
  \geq \frac{2}{\Nb}.
\end{equation}
Here $\Var{\hat{c}}$ is the variance of the estimate $\hat{c}$, and $\Nb$ is the mean number of binding events in time $T$.

For over 30 years,~\cref{eq:BergPurcell} was thought to constitute a fundamental physical limit on the accuracy of cellular chemosensation.
However, recent work focusing on highly specific receptor models~\cite{Mora2010,Endres2009a,Mehta2012, Lang2014a}  revealed this limit could be circumvented in two qualitatively distinct ways.
First, in the simple case of a two-state \object{}, an \obstraj{} (\IO) that has access to the entire \object{} trajectory of  binding and unbinding events, could outperform the \SO{} by performing maximum-likelihood estimation (MLE), obtaining an error $\epsilon_{\hat{c}}^{2} = \frac{1}{\Nb}$~\cite{Endres2009a}.
The \IO{} in this case outperforms the \SO{} by a factor of two, by employing the mean duration of unbound intervals, and ignoring the duration of bound intervals which contribute spurious noise because the transition rate out of the bound state is independent of $c$.
Second, even when the estimate is performed by a \SO{}, the Berg-Purcell limit can be overcome by energy-consuming non-equilibrium \objects{} with more than two states (\cref{fig:fig1}(d)), reflecting different \object{} conformations or phosphorylation states~\cite{Mora2010, Endres2009a, Lang2014a}.
Notably,~\cite{Lang2014a} numerically observed a tradeoff between error and energy for a very specific class of receptor models with states arranged in a ring.

While these more recent works demonstrate circumventions of the Berg-Purcell limit in highly specific models, they leave open foundational theoretical questions about the general interplay between estimation accuracy and \energy{} across the large space of possible complex non-equilibrium receptors.
Specifically, for a \SO, which may be implemented in a more biologically plausible manner than an  \IO, can we derive general \emph{analytic} bounds or exact formulas for accuracy in terms of energy expenditure for general classes of non-equilibrium \objects?
Can we exploit these formulas to find Pareto-optimal receptors through numerical optimization?

We address these questions by combining and extending stochastic thermodynamics~\cite{Seifert2008,Esposito2010,Sekimoto2010,Zhang2012,Ge2012,Seifert2012} and large deviation theory (LDT) of Markov chains~\cite{Touchette2009, Maes2008, Bertini2012, Bertini2014, Barato2015_ld, Barato2015, Gingrich2016b}.  
We derive a novel thermodynamic uncertainty relation connecting fluctuations in the time a stochastic process occupies a subset of states, and the energy dissipated by that process.
This relation is of independent interest to non-equilibrium statistical mechanics~\cite{Barato2015b, Barato2015, Gingrich2016b, Horowitz2020} and could have applications not only to cellular chemosensation, as discussed here, but also to understanding relations between energy and accuracy in other biological processes, like cellular motors and biological clocks~\cite{Li2002, Qian2007, Marsland2019}.

\paragraph*{Overall framework.}
A general non-equilibrium receptor can be modeled as a continuous-time Markov process~\cite{Horowitz2014,Lang2014a}, with $n$ different conformational or signaling states indexed by $i=0,\dots,n-1$.
The transition rate from state $i$ to $j$ is \(Q_{ij}\).
We assume some subset of these rates are binding transitions with rates proportional to concentration $c$ (see~\cite[\cref{sec:nonlinear}]{Supplemental} for a discussion of more complex dependence on concentration).
Over observation time $T$, the \object{} moves stochastically through a sequence of states 
yielding a random trajectory $\traj$. 
An \IO{} that has access to the entire trajectory (\cref{fig:fig1}(c)) can compute a MLE of the concentration $c$ via
\begin{equation}\label{eq:mledef}
  \hat{c} = \operatorname{argmax}_c  \log  \Prob\conds{\traj}{c} ,
\end{equation}
where $\Prob\conds{\traj}{c}$ denotes the probability distribution over \object{} state trajectories at a given concentration $c$.
To discuss the \SO, we further assume that binding transitions occur from a group of states defined as unbound, nonsignaling states, \(\CN \), to states we call the bound, signaling states, \(\CS \) (\cref{fig:fig1}(d)).
While an \IO{} can access the trajectory, we assume the \SO{} can only access the fraction of time spent in the signaling states, perhaps by counting the number of signaling molecules generated while the \object{} occupies those states. 
We note these assumptions are consistent with previous work~\cite{Endres2009a,Lang2014a}, though they exclude \objects{} with intermediate states that are bound but not signaling~\cite{Skoge2013}.

\paragraph*{\Object{} complexity and the \obstraj{}.}

We first ask if more states and transitions could yield improved accuracy relative to an \IO{} of a two state receptor~\cite{Endres2009a}.
 The properties of Markov processes~\cite[\cref{sec:traj}]{Supplemental}, imply the log probability of trajectory $\traj$ reduces to~\cite{Maes2008, Bertini2012, Bertini2014, Barato2015_ld}
\begin{equation}\label{eq:trajcollapseMain}
  \log \Prob\conds{\traj}{c}
  = -T \sum_{i \neq j} \brk{\pem_i Q_{ij} - \phiem_{ij} \log Q_{ij}},
\end{equation}
%
where \(\pem_{i} \) 
is the empirical density, or fraction of time the trajectory $\traj$ spends in state \(i\), and \(\phiem_{ij}\) is the empirical flux, or the number of transitions from state $i$ to state $j$ divided by $T$.

Maximizing \cref{eq:trajcollapseMain} w.r.t.\ $c$ yields the \IO{} estimate $\hat{c}$ in \cref{eq:mledef}.
When only transition rates from \(\CN \) to \(\CS \) are proportional to $c$ as in \cref{fig:fig1}(d),
\(
  \hat{c} = \frac{\Bindem}{\left. \Bindpem \right|_{\mathrlap{c = 1}}} 
\),
where $\Bindem \equiv \sumB \phiem_{ij}$ is the \object{}'s empirical binding rate along trajectory $\traj$, and $\Bindpem \equiv \sumB \pem_i Q_{ij}$, the expected binding rate conditioned on the empirical density $\pem_i$.
For two states, \(\hat{c}\) 
is inversely proportional to the total duration of unbound intervals, agreeing with~\cite{Endres2009a}.
However, this result goes beyond~\cite{Endres2009a} to reveal what function of the trajectory $\traj$ an optimal \IO{} must compute to estimate concentration for arbitrarily connected \objects{}, as in \cref{fig:fig1}(d).

The Cram\'er-Rao bound~\cite{Cover2006} lower bounds the \error{} $\epsilon_{\hat{c}}^{2}$ of the \IO{} through the Fisher information $J_{c}$ of the \object{} trajectory $\traj$ w.r.t. the external concentration $c$.
A simple calculation~\cite[\cref{sec:fisher}]{Supplemental} yields
\begin{equation}\label{eq:FisherGeneral}
  J_{c} = J_{c}^{0}
  + T\sum_{i\neq j}\pi_{i} Q_{ij} \: {\big[\partial_{c}\log{Q_{ij}}\big]}^2.
\end{equation}
Here \(\pi_i\) is the steady-state probability of state \(i\), and \(J_c^0 =  \sum_{i} \pi_{i}{(\partial_{c} \log{\pi_{i}})}^2\) is the Fisher information that the initial state contains about $c$.
The term linear in $T$ reflects additional information obtained from the entire trajectory.
Note only transition rates modulated by $c$ contribute information.
This result holds for arbitrary \objects{} as in \cref{fig:fig1}(c), but simplifies to \(J_{c} = J_{c}^{0} + \frac{T}{c^2} \Bindp\),
where  \(\Bindp = \sumB \pi_i Q_{ij}\) is the expected steady-state binding rate,
%
%
for \objects{} of the form in \cref{fig:fig1}(d) with transition rates linear in \(c\). 
Then the Cram\'er-Rao (CR) bound yields for large $T$,
\begin{equation}\label{eq:trajUncertainty}
  \epsilon_c^2 \equiv \frac{\Var{\hat{c}}}{c^2} \ge \frac{1}{J_c \,c^2} = \frac{1}{T\, \Bindp} = \frac{1}{\Nb}
\end{equation}
where \(\Nb\) is the expected number of binding events.
In~\cite{Supplemental} we directly calculate the variance of the \IO{} concentration estimate 
and demonstrate its error saturates the CR bound \cref{eq:trajUncertainty}.
This result generalizes~\cite{Endres2009a} from simple equilibrium two state \objects{} to arbitrarily connected nonequilibrium \objects{} of the form in \cref{fig:fig1}(d), confirming that any such energy-consuming nonequilibrium \object{} with binding rates proportional to concentration cannot outperform an \IO{} of a simple equilibrium two-state \object{}.

\paragraph*{Fluctuations and gain determine \obsq{} performance.}

The \IO{} estimate in \cref{eq:mledef} 
requires computing a complex function of
the \object{} trajectory $\traj$, which may not be biologically plausible.
We therefore explore a \SO, which estimates concentration using only the fraction of time
the \object{} is bound (signaling), denoted by \(\qem = \sum_{i \in \CS} \pem_i\).
Due to randomness in $\traj$, $\qem$ fluctuates about its mean \(\qp = \sum_{i \in \CS} \pi_i\), which depends on the concentration $c$. 
Given the observable $\qem$, one can then estimate $\hat{c}$ by solving \(\qp(\hat{c}) = \qem\).
Standard error propagation then yields,
%
\begin{equation}\label{eq:stderror}
  \epsilon_{\hat{c}}^2
  = \frac{\Var{\hat{c}}}{c^2}
  = \brk{c\, \diff{\qp}{c}}^{\mathrlap{-2}} \Var{\qem}
  .
\end{equation}
Thus a larger variance $\Var{\qem}$ in the time spent bound increases the error $\epsilon_{\hat{c}}^2$, while a larger gain $\left|\diff{\qp}{c}\right|$   decreases it.
We next compute and bound this variance and gain.

\paragraph*{A thermodynamic uncertainty relation for densities.}
We first derive a lower bound on the variance $\Var{\qem}$ using stochastic thermodynamics and LDT~\cite{Touchette2009} of Markov processes.
A random trajectory $\traj$ of duration $T$ in a general Markov process will yield an empirical density $p_{i}^{T}$ and an empirical current \(\jem_{ij} = \phiem_{ij} - \phiem_{ji}\), which corresponds to the \emph{net} number of transitions from $i$ to $j$ divided by $T$.
As $T\rightarrow\infty$, these random variables will converge to their mean values, corresponding to the steady state probabilities \(\pi_{i} = \lim_{T \rightarrow \infty} \pem_{i} \) and steady state currents
\(\jp_{ij} \equiv \pi_{i}Q_{ij} - \pi_{j}Q_{ji} = \lim_{T \rightarrow \infty} \jem_{ij} \).
At large but finite $T$, \(\pem\) and \(\jem\) fluctuate about their means, and their joint distribution takes the form  \(\Prob(\pem = p, \jem = j) \propto \e^{-T I(p,j)}\)~\cite{Maes2008, Bertini2012, Bertini2014, Barato2015_ld, Supplemental}.
Here  \(I(p,j)\) is a large deviation rate function that achieves its minimum at $p = \pi$ and $j = \jp$, and describes how fluctuations in \(\pem\) and \(\jem\) are suppressed.
This rate function is 
 $I(p,j) = \sum_{i\,<\,j} j_{ij}(\arcsinh \frac{j_{ij}}{a_{ij}} - 
            \arcsinh \frac{j_{ij}^{p}}{a_{ij}}) 
    - \big(\sqrt{a_{ij}^{2} + j_{ij}^{2}}  -
             \sqrt{a_{ij}^{2} + j_{ij}^{p\,2}}\big)$,
where \(a_{ij} \equiv 2\sqrt{p_{i}p_{j}Q_{ij}Q_{ji}} \) and \(j^{p}_{ij} \equiv p_{i}Q_{ij} - p_{j}Q_{ji}\)~\cite{Maes2008,Barato2015_ld,Bertini2012,Bertini2014}.


Similarly, at large but finite $T$, the distribution of the fraction of time time spent bound, namely \(\qem = \sum_{i \in \CS} p_{i}\), takes the form \(\Prob(\qem = q) \propto \e^{-T I(q)}\).
Here the large deviation rate function $I(q)$ achieves its minimum at the mean value $\qp \equiv \sum_{i \in \CS} \pi_i$, and describes how deviations in \(\qem\) from its mean are suppressed.
The variance of  \(\qem\) is given by \(1/(TI''(\qp))\)~\cite{Touchette2009}, so any upper bound on \(I''(\qp)\) will yield a lower bound on the variance of \(\qem\).

One can obtain \(I(q)\) from the more general rate function \(I(p,j)\) through the contraction principle~\cite{Touchette2009}, which states that \(I(q) = \inf_{p,j}
I(p,j)\), subject to the constraints \(\sum_j j_{ij} = 0\: \forall \, i\), \(\sum_i p_i = 1\) and \(\sum_{i \in \CS} p_i = q\).
Instead of calculating this directly, the infimum can be bounded by evaluating \(I(p,j)\) for a choice of \(j = j^*(q)\) and \(p = p^*(q)\) satisfying the same constraints.
With the following choice of \(p^*(q)\) and \(j^*(q)\),
\begin{equation}\label{eq:contraction}
\begin{gathered}
  I(q) \le I(p^*,j^*),
  \quad
p^*_{i}(q) =
   \begin{cases}
      \frac{q}{\qp} \pi_{i} & i \in \CS, \\
      \frac{1-q}{\qm} \pi_{i} & i \in \CN,
   \end{cases}\\
j^*_{ij}(q) = \brk{\frac{q(1-q) + \qpm}{2\qpm}} \jp_{ij}. 
\end{gathered}
\end{equation}
This is a simple choice that satisfies the constraints mentioned above, as well as \(j^*_{ij}(\qp) = \jp_{ij}\) and \(p^*_i(\qp) = \pi_i\), ensuring that the inequality in~\cref{eq:contraction} is saturated at the minimum, \(q = \qp \).  
The coefficient in brackets in \cref{eq:contraction} is chosen to maximize the tightness of the eventual bound. 

Following the approach of~\cite{Gingrich2016b}, inserting our choice of \(p^{*}, j^*\) into $I(p,j)$ leads to an explicit upper bound on \(I''(\qp)\) in terms of the total \energy{} rate of the receptor (in units of \(\kt \)), defined as \(\Sigmap = \sum_{i < j} \jp_{ij} \log \frac{\phip_{ij}}{\phip_{ji}}\)~\cite{Seifert2008}. 
We then find a lower bound on the variance of \(q\)~\cite[\cref{sec:ldt_bnd}]{Supplemental}:
\begin{equation}\label{eq:epsqbound}
  \Var{q} \left[T \Sigmap + 4 \Nb \right]  \,  \ge \, 8\, \qpmb^{2}.
\end{equation}
\Cref{eq:epsqbound} can be thought of as a new, general thermodynamic uncertainty relation which implies that the more energy $T \Sigmap$ a system consumes, the more reliable the occupation time for a pool of states can become.
This can be compared to another thermodynamic uncertainty relation connecting increased \energy{} to a reduction in current fluctuations in general stochastic processes~\cite{Gingrich2016b,Barato2015b}.
Our result in \cref{eq:epsqbound} adds pooled state occupancies to the class of observables for which thermodynamic uncertainty relations can be generally proven.

\paragraph*{An energy-accuracy tradeoff for the \obsq{}.}
The gain $c\, \diff{\qp}{c}$ in \cref{eq:stderror} can be calculated for arbitrary nonequilibrium processes using the known relationship between first passage times and the sensitivity of Markov chain stationary distributions~\cite{Supplemental,cho2000markov}.
Our formulae in~\cite[\cref{sec:jacobian}]{Supplemental} simplify to $c\, \diff{\qp}{c} = \qpm$ for nonequilibrium \objects{} of the form in \cref{fig:fig1}(d) with only one nonsignaling state and binding transitions linear in \(c\).\footnote{The assumptions of only one nonsignaling state and binding transitions that are linear in concentration are required at this stage of the derivation only.  
If the binding transition rates are related to concentration through a power law, the conclusions are similar, see~\cite[\cref{sec:nonlinear}]{Supplemental}.}
Inserting this result for gain and the relation for variance in \cref{eq:epsqbound} into \cref{eq:stderror}, yields a general lower LDT-bound on error in terms of energy $T \Sigmap$ and mean binding events \(\Nb\):
\begin{equation}\label{eq:finalResult}
  \epsilon_{\hat{c}}^{2} \,
  \ge \, \frac{8}{T\Sigmap + 4\Nb}.
\end{equation}
This recovers Berg-Purcell~\cref{eq:BergPurcell} at zero energy rate $\Sigmap$.

\begin{figure}[th]
  \includegraphics[width=1.0\linewidth]{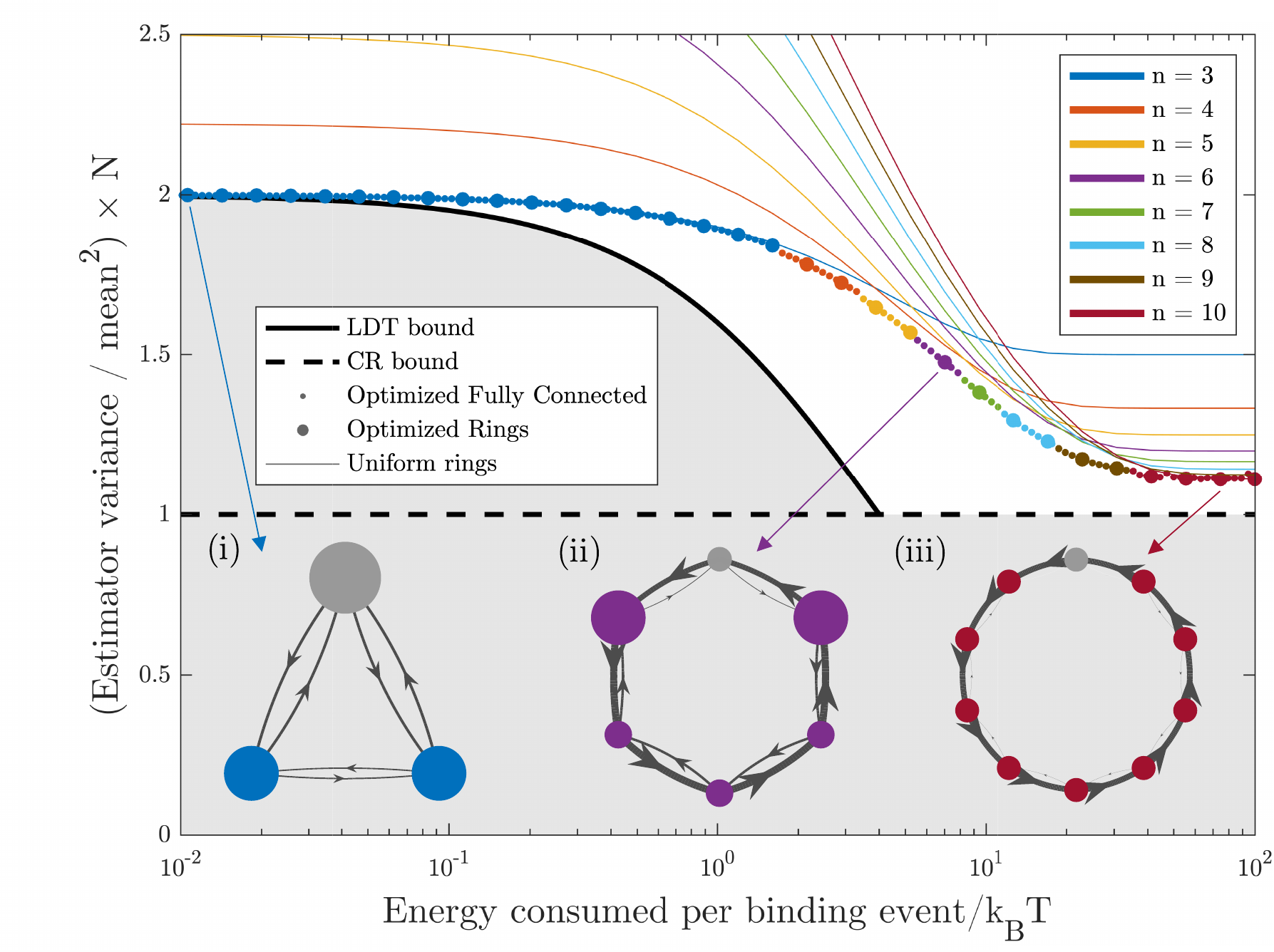}
  \caption[Numerically optimized \objects{}]
  {\label{fig:fig2} Energy-accuracy tradeoffs. The LDT bound \cref{eq:finalResult} (solid black), and the CR bound \cref{eq:trajUncertainty} (dashed black) together yield a forbidden region of achievable error and energy (gray). 
  Solid circles show the minimal error achieved by fully connected (small circles) and ring (large circles) \objects{} after numerically minimizing~\cref{eq:exact} (see~\cite[\cref{sec:conc}]{Supplemental}) w.r.t all transition rates, with an energy constraint.
    Data point color reflects the error of the smallest \object{} that is within 1\% of the \error{} of the best performing receptor obtained at each energy, indicating the smallest number of states for which adding states (up to 10) would not significantly lower error. 
     Thin lines show the performance of $n$-state ring receptors with uniform transition rates in each direction. 
    (i-iii) Three optimal \objects{} found at different \energy{} levels indicated by arrows.
    Colored (gray) nodes in the diagrams represent signaling (nonsignaling) states.
    Node radii are proportional to steady state probabilities $\pi_i$ and edge widths are proportional to steady state fluxes.
  }
\end{figure}

Overall, \cref{eq:finalResult} is a generalization of the Berg-Purcell limit to general energy consuming non-equilibrium \objects{} of the form in \cref{fig:fig1}(d), with one nonsignaling (unbound) state, but an arbitrary network of signaling (bound) states.  
This LDT-bound is clearly not tight as \( \Sigmap \rightarrow \infty \), but for finite \( \Sigmap \) it provides a simple energy-based bound on error that is independent of both the detailed number and connectivity of receptor states. 
At \( \Sigmap \geq 4\Nb/T\) the CR-bound in \cref{eq:trajUncertainty} becomes more stringent than the LDT-bound in \cref{eq:finalResult}, and any bound on the \IO{} must also apply to the \SO.  
Thus the combined bound (the maximum of the CR and LDT bounds) yields a forbidden region of error versus energy  (\cref{fig:fig2}).

\paragraph*{Exact estimation error for the \obsq{}.}

\Cref{eq:finalResult} provides a lower bound on the \error{} because our choice of \(p^*, j^*\) in \cref{eq:contraction} does not achieve the infimum of the contraction.
When the contraction of the rate function \(I(p,j)\) is expanded as a Taylor series in \((q-\qp)\),
one can compute the optimal \(p\) and \(j\) to leading order, which allows us to find the second derivative term in the Taylor expansion of \(I\) and hence the uncertainty \(\epsilon_{\hat{c}}^{2}\) exactly \cite[\cref{sec:varc}]{Supplemental}.
Under the same assumption of one nonsignaling state, we find
\begin{equation}\label{eq:exact}
  \epsilon_{\hat{c}}^{2} = \frac{2}{\Nb} \frac{\tunbind}{\thold} ,
\end{equation}
where \(\tunbind\) is the mean time until the next unbinding event given the \object{} is in a signaling (bound) state, \(\tunbind =\sum_{i \in \CS} \mathbf{T}_{i\CN} \Prob \cond{x(t) = i}{x(t) \text{ in } \CS}\),
\(\thold\) is the mean duration of a full journey through the signaling states, \(\thold = \sum_{i \in \CS} \mathbf{T}_{i\CN}\, \Prob \cond{x(t) = i}{x(t) \text{ just entered } \CS}\),
and \(\mathbf{T}_{i\CN}\) is the mean first passage time from state \(i\) to the single nonsignaling state.
In~\cite[\cref{sec:varc}]{Supplemental} we have numerically verified this formula by comparing it with Monte-Carlo simulations.
For a two-state system, \(\tunbind = \thold\) because the unbinding process is memoryless, and \cref{eq:exact} reduces to the Berg-Purcell limit in \cref{eq:BergPurcell}.
Thus \cref{eq:exact} is another generalization of Berg-Purcell to general energy consuming non-equilibrium \objects{} of the form in \cref{fig:fig1}(d), with one nonsignaling state, but an arbitrary network of signaling states.
While \cref{eq:exact}, and its generalization to multiple nonsignaling states~\cite[\cref{sec:varc}]{Supplemental}, gives an exact formula for error that allows us to search for Pareto-optimal receptors, our LDT-bound in \cref{eq:finalResult} makes manifest a connection between error and energy.


%

\paragraph*{Pareto-optimal error reduction via energy consumption requires more states and is achievable by rings.} 
\Cref{fig:fig2} compares the error bounds in \cref{eq:trajUncertainty,eq:finalResult} to numerical minimization of \cref{eq:exact} at fixed energy consumption for receptors of increasing size.
The union of the CR and LDT lower bounds is respected by all models found.
For a given energy consumption and number of states $n$,
we minimized error over all possible fully connected receptors for every partition of $n$ into signaling and nonsignaling states.
We observed that for all receptor sizes and energies studied, the error achieved with a single nonsignaling state is not outperformed by any other partition 
\cite[\cref{sec:varyN}]{Supplemental}, so we focus on this case.
To explore the role of the number of states in the Pareto-optimal tradeoff between energy and accuracy, in \cref{fig:fig2} we numerically minimized error over all possible fully connected \objects{}, including the number of states (up to 10), for a given energy consumption. 
We found that as energy increases, near minimal error is achievable only by increasing the number of states (\cite[\cref{fig:fixedn}]{Supplemental}). 
\Cref{fig:fig2}~(i--iii) depict optimal minimal size \objects{} obtained at different levels of \energy{}.
At low \energy{} the LDT bound is tight, and the optimal \object{} is equivalent to a two-state \object{} with the signaling states behaving like one coarse-grained state (see~\cite[\cref{sec:optNetworks}]{Supplemental}), and there is no advantage to adding $> 3$ states.
The LDT-bound \cref{eq:finalResult} becomes increasingly loose as energy consumption increases and the minimal achievable error of optimal \objects{} at fixed $n$ saturates at a level that depends on the number of states. 
At high \energy{}, the optimal \object{} approaches a many-state uniform ring with more asymmetric transition rates at higher energies.
As the number of states \(n \rightarrow \infty\), the minimal achievable error at high \energy{} saturates the CR-bound \cref{eq:trajUncertainty}.
Thus the combined LDT and CR bound is tight at both high and low energy. 
We repeated the optimization with \objects{} restricted to ring topologies with arbitrary transition rates, and found that the best possible rings perform indistinguishably from fully connected networks at every energy level (\cref{fig:fig2}).
This suggests that the Pareto optimal tradeoff between energy and accuracy is achieved by non-uniform ring networks of increasing size and energy consumption.


\paragraph*{Uniform ring receptors are not Pareto-optimal.}
For ring receptors with uniform transition rates in each direction, we can analytically compute $\tunbind$, $\thold$, and therefore $\epsilon_{\hat{c}}^{2}$ in \cref{eq:exact}, as a function of $n$, $\Sigmap$, and $\Nb$~\cite[\cref{sec:uniring}]{Supplemental}, obtaining 
\begin{equation}\label{eq:uniform}
  \epsilon_{\hat{c}}^{2}
    = \frac{n \coth\brk{\frac{\sigma}{2n}} 
              \prn{n \coth\brk{\frac{\sigma}{2}} - \coth\brk{\frac{\sigma}{2n}}}}
           {\overline{N}{(n-1)}^2},
\end{equation}
where $\sigma$ is defined via \( \frac{\Sigmap}{\Bindp} = \sigma \tanh \brk{\frac{\sigma}{2n}} \).
This error versus energy for $n=3,\dots,10$ is plotted in \cref{fig:fig2}. 
The lower envelope of these curves, corresponding to minimizing \cref{eq:uniform} w.r.t $n$ at fixed energy per binding event \( \frac{\Sigmap}{\Bindp} \), yields an upper bound on the numerically derived Pareto optimal tradeoff. 
As seen in \cref{fig:fig2}, at intermediate energy consumption, this upper bound is slightly higher than the minimal error achieved by nonuniform rings, indicating nonuniform transition rates are required for ring receptors to be optimal at intermediate energy. 
However, as $\Sigmap\rightarrow\infty$, $\epsilon_{\hat{c}}^{2} \rightarrow \frac{1}{\Nb}\frac{1}{1-1/n}$, indicating that a \SO{} of uniform rings can approach the CR bound at large energy and $n$. 
The optimality of single path rings is reminiscent of the optimality of single path chains for hitting times, observed numerically in~\cite{Escola2009} for a different notion of energy.
However, large uniform rings can be highly suboptimal under a \SO{} at low energy; as $\Sigmap\rightarrow 0$, $\epsilon_{\hat{c}}^{2} \rightarrow \frac{2}{\Nb}\frac{n(n+1)}{6(n-1)}$.
This 
reproduces Berg-Purcell in~\cref{eq:BergPurcell} for $n=2,3$, but is much worse for $n>3$ (\cref{fig:fig2}).
Thus larger receptors \textit{must} consume more energy to be optimal. 

\paragraph*{Discussion.}
We derived several general results 
\cref{eq:FisherGeneral,eq:trajUncertainty,eq:finalResult,eq:exact} 
delineating fundamental performance limits of cellular chemosensation using arbitrarily complex energy consuming nonequilibrium receptors, as a joint function of observation time, \energy{} rate, number of states, and the sophistication of the downstream observer.
Along the way we have also derived a general thermodynamic uncertainty relation \cref{eq:epsqbound} which reveals one must pay a universal energetic cost for reliable occupation time in any physical process.
We hope these analytic relations between time, energy and accuracy will find further applications in myriad biological and physical processes~\cite{Qian2007,Lan2012a, Mehta2016, Parrondo2015a, Lahiri2016, Still2012,Boyd2017,Marzen2020}.

\bibliography{journals, CellSensePRL}

%
\let\oldaddcontentsline\addcontentsline
\renewcommand{\addcontentsline}[3]{}
\let\addcontentsline\oldaddcontentsline

\clearpage
\onecolumngrid
\newgeometry{margin=1.25in}
\begingroup
  \makeatletter
    \renewcommand{\footnotesize}{%
      \@setfontsize\footnotesize\@xpt{12pt}%
      \abovedisplayskip 10\p@ \@plus2\p@ \@minus5\p@
      \belowdisplayskip \abovedisplayskip
      \abovedisplayshortskip \z@ \@plus3\p@
      \belowdisplayshortskip 6\p@ \@plus3\p@ \@minus3\p@
      \def\@listi{%
        \leftmargin\leftmargini
        \topsep 6\p@ \@plus2\p@ \@minus2\p@
        \parsep 3\p@ \@plus2\p@ \@minus\p@
        \itemsep \parsep
      }%
    }%
    \onecolumn@grid@setup
    \renewcommand{\set@footnotewidth}{\set@footnotewidth@one}
  \makeatother
  \fontsize{12}{14}\selectfont

\part*{Supplementary Information}

\vspace{1cm}
\tableofcontents


\newpage
\section{Overview}\label{sec:overview}

In this supplement we provide complete derivations of several results presented in the main text, as well as background material on Markov processes and large deviation theory for a general physics audience.

In \cref{sec:idealobs} below we derive the maximum-likelihood estimate (MLE) and \cref{eq:FisherGeneral} of the main text.
The MLE reveals the computation the \obstraj{} (\IO{}) must make to construct an estimate of the concentration from knowledge of the entire trajectory of \object{} states over a given time interval.
Correspondingly, \cref{eq:FisherGeneral} describes the Fisher information that the entire \object{} state trajectory contains about the external concentration.
The reciprocal of this Fisher information bounds the error of the \IO{} estimate through the Cram\'er-Rao bound.

In \cref{sec:ldt_bnd,sec:jacobian} we derive the main text's \cref{eq:epsqbound,eq:finalResult}.
\Cref{eq:epsqbound} describes a thermodynamic uncertainty relation revealing that energy must be spent to reduce fluctuations in the time a physical process spends in a subset of states.
\Cref{eq:finalResult} describes how this thermodynamic uncertainty relation, when combined with a computation of the \object{} gain, yields a lower bound on estimation error in terms of energy consumption.

In \cref{sec:conc} we use large deviation theory to derive exact formulae for the \error{} for the \obstraj{} (\IO) and the \obsq{} (\SO), including the special case of a uniform ring \object{}.
The second formula (main text \cref{eq:exact}) was used for the numerical optimization of \SO{} performance over the space of \objects{} in \cref{fig:fig2} of the main text.

In \cref{sec:numerics} we provide details for the numerical computations in the main text \cref{fig:fig2}.

Finally, to make this supplement self contained, we provide \cref{sec:markovprimer,sec:ldt_review} with brief reviews of the theory of continuous-time Markov processes and the large deviation theory for empirical density, flux and current.


\section{Information and estimation accuracy for the \obstraj{}}\label{sec:idealobs}

\subsection{Fisher information in a Markovian trajectory}\label{sec:fisher}

In this \namecref{sec:fisher} we derive the Fisher information from an extended observation of a system with Markovian dynamics, \cref{eq:FisherGeneral} in the main text.
We first consider a discrete time Markov process, and will later take the limit as the size of the discrete time steps \(\Delta t\) become vanishingly small.

The discrete-time transition matrix is given by \(M\), where \(M_{ij}\) is the probability of transition from state \(i\) to state \(j\) if the system is in state \(i\) at a particular time step.
For a set of states labeled by \(i\), we define \(\pi_{i}\) as the steady state distribution, which satisfies \(\eq \Md = \eq \) and has elements which sum to 1.
The matrix \(\Md \) can be expanded in terms of the continuous time transition rate matrix \(\MM \), which has elements \(\MMm_{ij}\) and obeys \(\sum_{j}\MMm_{ij} = 0\) (see \cref{eq:mastermatrix} in \cref{sec:mastereqn}).
\begin{equation}
\Md = \e^{\MM\Delta t} = \I + \MM \Delta t + \ldots
\end{equation}

We would like to consider the general probability of a Markovian trajectory from a state \(x_0\) at \(t = 0\) to state \(x_n\) at \(t = n\Delta t\).%
\footnote{
In the main text we described a trajectory by the sequence of states visited, \(\brc{x_0, \ldots, x_m}\) and the transition times \(\brc{t_0, \ldots, t_m}\).
For \emph{this section only}, we find it more convenient to describe the trajectory by the identity of the state occupied at each of a discrete set of time steps, \(x_k = x(k\Delta t), k = 0, \ldots, n\).
In the continuum limit, \(\Delta t \rightarrow 0, n \rightarrow \infty \), the two descriptions contain the same information.
}
Assuming that the system begins in the steady state distribution, the probability of this trajectory in discrete time is given by \(\Prob\event{x_0,\ldots,x_n} = \pi_{x_{0}} M_{x_{0}x_{1}}M_{x_{1}x_{2}} \ldots M_{x_{n-1}x_{n}}\).
We can now directly calculate the Fisher information matrix for this distribution with respect to the parameter \(\lambda_\mu \), using the notation \(\partial_\mu \equiv \frac{\partial}{\partial \lambda_\mu}\):
\begin{equation}\label{FisherMarkov}
\begin{aligned}
  J_{\mu\nu} = \smashoperator{\sum_{x_0,\dots,x_n}} \pi_{x_{0}}
         M_{x_{0}x_{1}} \dots M_{x_{n-1}x_{n}}
         &\Big[ \partial_{\mu} \log(  \pi_{x_{0}} M_{x_{0}x_{1}}\dots M_{x_{n-1}x_{n}} ) \Big] \\
         \times &\Big[ \partial_{\nu} \log(  \pi_{x_{0}} M_{x_{1}x_{2}} \dots M_{x_{n-1}x_{n}} ) \Big].
\end{aligned}
\end{equation}

We now recognize that the Fisher information matrix \cref{FisherMarkov} can be rewritten as (using the fact \(\sum_{k} M_{jk} = 1\) and \(\sum_j \pi_j M_{jk} = \pi_k\))
\begin{equation}
\begin{alignedat}{2}
  J_{\mu\nu} =
    &\sum_{x_0} \pi_{x_{0}} \partial_{\mu} \log{\pi_{x_0}} \partial_{\nu} \log{\pi_{x_0}} \\
        &+   \smashoperator{\sum_{x_0 \dots x_n}} \pi_{x_{0}} M_{x_{0}x_{1}}M_{x_{1}x_{2}} \dots &M_{x_{n-1}x_{n}}
        & \Big[ \partial_{\mu} \big(\log{M_{x_{0}x_{1}}} + \cdots + \log{M_{x_{n-1}x_{n}}} \big) \Big] \\
 &&\times & \Big[ \partial_{\nu} \big( \log{ M_{x_{0}x_{1}}} + \cdots + \log{M_{x_{n-1}x_{n}}} \big) \Big].
\end{alignedat}
\end{equation}
or,
\begin{equation}\label{Fisher2}
J_{\mu\nu} =  J_{\mu \nu}^{0}
    + \smashoperator{\sum_{x_0 \dots x_n}} \pi_{x_{0}} M_{x_{0}x_{1}}M_{x_{1}x_{2}} \cdots M_{x_{n-1}x_{n}}
      \Big[ \partial_{\mu} \sum_{k = 0}^{n-1} \log{M_{x_{k}x_{k+1}}} \Big]
      \Big[ \partial_{\nu} \sum_{j = 0}^{n-1}  \log{ M_{x_{j}x_{j+1}}} \Big]
\end{equation}
where \(J_{\mu \nu}^{0}\) is the Fisher information matrix for a random variable representing a single observation of the system state, and the indices \(k\) and \(j\) index the time steps in the measurement interval.
The expression \cref{Fisher2} simplifies if we write the sums inside the brackets on the right hand side out term-by-term.
For \(k = 0, \,j = 0\), the second term on the right hand side simplifies to
\begin{multline}\label{keqj}
 \smashoperator{\sum_{x_0 \dots x_n}} \pi_{x_{0}} M_{x_{0}x_{1}} \dots M_{x_{n-1}x_{n}}
   \Big[ \partial_{\mu} \log{M_{x_{0}x_{1}}} \Big]
   \Big[ \partial_{\nu} \log{ M_{x_{0}x_{1}}} \Big]
   \\
   =  \smashoperator{\sum_{x_0, x_1}} \pi_{x_{0}} M_{x_{0}x_{1}}
   \Big[ \partial_{\mu} \log{M_{x_{0}x_{1}}} \Big]
   \Big[ \partial_{\nu} \log{M_{x_{0}x_{1}}} \Big].
\end{multline}
Similarly, for \(k = 0,\, j = 1\), we have
\begin{equation}\label{kneqj}
\begin{split}
 \smashoperator{\sum_{x_0 \dots x_n}} \pi_{x_{0}} & M_{x_{0}x_{1}} \dots M_{x_{n-1}x_{n}}
    \Big[ \partial_{\mu}  \log{M_{x_{0}x_{1}}} \Big]
    \Big[ \partial_{\nu}   \log{ M_{x_{1}x_{2}}} \Big] \\
 = & \smashoperator{\sum_{x_0, x_1, x_2}} \pi_{x_{0}} M_{x_{0}x_{1}} M_{x_{1}x_{2}}
    \Big[ \partial_{\mu}  \log{M_{x_{0}x_{1}}} \Big]
    \Big[ \partial_{\nu}   \log{ M_{x_{1}x_{2}}} \Big] \\
 =  & \smashoperator{\sum_{x_0, x_1, x_2}} \pi_{x_{0}}  \partial_{\mu}  M_{x_{0}x_{1}}\partial_{\nu}   M_{x_{1}x_{2}} = 0.
\end{split}
\end{equation}
In the same fashion, all \(k = j\) terms in \cref{Fisher2} give an expression similar to \cref{keqj} and all \(k \neq j\) terms vanish as in \cref{kneqj}.
Our expression for the Fisher information matrix then becomes:
\begin{equation}
\begin{split}
  J_{\mu\nu} =  J_{\mu \nu}^{0} + \sum_{k = 1}^{n-1}
    \smashoperator[r]{\sum_{x_k,x_{k+1}}} \pi_{x_k} M_{x_k,x_{k+1}}\,
    \partial_{\mu}\log{M_{x_k,x_{k+1}}}\, \partial_{\nu}\log{M_{x_k,x_{k+1}}}.
\end{split}
\end{equation}
Given that \(M\) is not changing in time, after relabeling \(x_k \rightarrow i\) and \(x_{k+1} \rightarrow j\) all terms in the sum over \(k\) are identical.
We therefore find:
\begin{equation}\label{FisherDiscrete}
  J_{\mu\nu} =  J_{\mu \nu}^{0} + n\sum_{i,j}\pi_{i} M_{ij}\,
  \partial_{\mu}\log{M_{ij}}\, \partial_{\nu}\log{M_{ij}}.
\end{equation}
Lastly, we take the continuous time limit by sending \(\Delta t \rightarrow 0\).
For infinitesimal \(\Delta t\),  \(M_{ij} = \MMm_{ij}\Delta t\) for \(i \neq j\) and \(M_{ii} = 1 + \MMm_{ii}\Delta t\).
We can then rewrite \cref{FisherDiscrete} as
\begin{equation}\label{takinglimit}
\begin{split}
 J_{\mu\nu} & = J_{\mu \nu}^{0} + n\sum_{i\neq j}\pi_{i} \MMm_{ij}\Delta t \: \partial_{\mu}\log(\MMm_{ij}\Delta t) \partial_{\nu}\log(\MMm_{ij}\Delta t) \\
 & +   n\sum_{i=j}\pi_{i} (1 + \MMm_{ii}\Delta t) \: \partial_{\mu}\log(1 + \MMm_{ii}\Delta t)\partial_{\nu}\log{(1 + \MMm_{ii}\Delta t)}.
 \end{split}
\end{equation}
In the limit \(\Delta t \rightarrow 0\), \(\Delta t \log{\Delta t} \rightarrow 0\), and
\(\log{(1 + \MMm_{ii}\Delta t)} \approx \MMm_{ii} \Delta t\).
Defining \(T \equiv n \Delta t\), we therefore find
\begin{equation}\label{s:FisherResult1}
J_{\mu\nu} = J_{\mu \nu}^{0} +   T\sum_{i\neq j}\pi_{i} \MMm_{ij} \: \partial_{\mu}(\log{\MMm_{ij}})\partial_{\nu}(\log{\MMm_{ij}}),
\end{equation}
where we have recognized that the \(i = j\) terms from \cref{takinglimit} all vanish in the limit \(\Delta t \rightarrow 0\).

We then assume that the signal to be estimated is a scalar denoted by \(c\), which could represent an external concentration of some ligand.
For a scalar parameter, the Fisher information of the entire trajectory then becomes:
\begin{equation}\label{FisherScalar}
J_{c} = J_{c}^{0} +   T\sum_{i\neq j}\pi_{i} \MMm_{ij} \: \brk{\partial_{c}\log{\MMm_{ij}}}^2,
\end{equation}
which is \cref{eq:FisherGeneral} in the main text.

If we specialize to the models of receptors studied in the main text,
the only off-diagonal transition rates that depend on \(c\) are those along the edges in the set \(\CBj\):
the edges that start in the set \(\CN\) and end in \(\CS\).
As those transition rates are proportional to \(c\), \cref{FisherScalar} reduces to:
\begin{equation}\label{eq:FisherConc}
  J_c = J_c^0 + T \sum_{ij \in \CBj} \frac{\pi_i \MMm_{ij}}{c^2}
      = J_c^0 + \frac{T \Bindp}{c^2}.
\end{equation}
%
This leads to a lower bound on the uncertainty of any unbiased estimate of \(c\),
via the Cram\'er-Rao bound~\cite{Cramer1945,Rao1945}.


\subsection{Maximum likelihood estimation for the \obstraj{}}\label{sec:mle}

In the previous \namecref{sec:fisher} we computed the Fisher information for the \obstraj{},
which leads to a lower bound on the uncertainty of any estimate of \(c\).
In general, the maximum likelihood estimator saturates the Cram\'er-Rao bound asymptotically,
in the limit of a large number of independent observations~\cite{Haldane1956}.
We compute this estimator in this \namecref{sec:mle}.
We will postpone calculating its variance to \cref{sec:mlevar}.

In \cref{sec:traj} we see that, when the duration of observation is large,
the likelihood of any single trajectory collapses to a function of certain summary statistics:
the empirical density, \(\pem_i\), the fraction of time spent in state \(i\),
and the empirical flux, \(\phiem_{ij}\), the rate at which transitions from state \(i\) to \(j\) occur (see \cref{sec:empirical} for precise definitions).
In \cref{eq:ldtregroup} we see that the likelihood is:
\begin{equation}\label{eq:trajcollapse}
  \log \Prob\cond{x(t)}{c}
    = -T \sum_{i \neq j} \brk{\pem_i \MMm_{ij} - \phiem_{ij} \log \MMm_{ij}},
\end{equation}
where \(\MMm_{ij}\) is the source of dependence on \(c\).

If we use the notation \(\phipem_{ij} = \pem_i \MMm_{ij}\), the maximum of this function must satisfy
\begin{equation}\label{eq:mlegen}
  \pdiff{\log\Prob\cond{x(t)}{c}}{c}
    = T \sum_{i \neq j} \prn{\phiem_{ij} - \phipem_{ij}} \pdiff{\log \MMm_{ij}}{c} = 0.
\end{equation}
Now we can specialize to the models of receptors studied in the main text,
where the only off-diagonal transition rates that depend on \(c\) are those along the edges in the set \(\CBj\).
As those transition rates are proportional to \(c\), \cref{FisherScalar} reduces to:
\begin{equation}
  \pdiff{\log\Prob\cond{x(t)}{c}}{c}
    = \frac{T}{c}\prn{\Bindem - \Bindpem} = 0.
\end{equation}
Because \(\Bindpem\) is proportional to \(c\), the maximum likelihood estimator is
\begin{equation}\label{eq:mleconc}
  \hat{c} = \frac{\Bindem}{\left. \Bindpem \right|_{\mathrlap{c = 1}}}.
\end{equation}
We will compute the variance of this estimator for large \(T\) in \cref{sec:mlevar}.


\section{A thermodynamic uncertainty principle for density}\label{sec:ldt_bnd}

Here we present a derivation of \cref{eq:epsqbound} in the main text, which constitutes a thermodynamic uncertainty relation connecting fluctuations in the fraction of time a physical process spends in a pool of states to the energy consumption rate of that process.
This uncertainty relation reveals that one cannot reduce fluctuations in total occupation time without paying an energy cost.

We make use of a known result that the empirical density and currents for continuous-time Markov processes obey a large deviation principle with a known joint rate function.
The large deviation rate function \(I(p,j)\) describes both fluctuations in the empirical quantities \(p\) and \(j\) around their steady states and highly unlikely large deviations~\cite{Touchette2009}.
This rate function is known to take the following form~\cite{Barato2015} (see also \cref{sec:currentLDT}):
\begin{equation}
I(p,j) = \sum_{i\,<\,j} \Psi(j_{ij},j^{p}_{ij},a^{p}_{ij})
\end{equation}
with (dropping the state indices \(i,j\) for notational simplicity)
\begin{equation}\label{PsiDef2}
\begin{aligned}
\Psi(j, j^{p}, a^{p}) = j\Big(\arcsinh \frac{j}{a^p} - \arcsinh \frac{j^{p}}{a^p}\Big)
 - \Big(\sqrt{a^{p \,2} + j^{2}} - \sqrt{a^{p \,2} + j^{p\,2}}\Big)
\end{aligned}
\end{equation}
where  \(a^{p}_{ij} \equiv 2\sqrt{p_{i}p_{j}\MMm_{ij}\MMm_{ji}} \) and \(j^{p}_{ij} \equiv p_{i}\MMm_{ij} - p_{j}\MMm_{ji}\).
We also require that the probability current is conserved, \(\sum_j j_{ij} = 0\) for all nodes indexed by \(i\).

For the purposes of this sensing problem, we are interested in the rate function of the density in a subset of states we call the signaling states, \(q = \sum_{i \in \mathcal{S}}  p_i\).
In the main text, we argued that we can bound this rate function by repeated application of the contraction principle, such that
\begin{equation}\label{Iqbound2}
  I(q) \le I_{b}(q) = \sum_{i\,<\,j} \Psi(j^*_{ij}, j^{p^*}_{ij}, a^{p^*}_{ij}).
\end{equation} is given by
where \(j^*\) and \(p^{*}\) are arbitrary choices for \(j\) and \(p\) in place of evaluating the infimum.

As discussed in the main text, we are interested in the variance of the \sigfrac{} \(q\), which is given by \(1/(TI''(\qp))\)~\cite{Touchette2009}.
Therefore, we are interested in bounding the quantity \(I''(\qp)\),
\begin{equation}\label{secondDbound2}
I''(\qp) \le I_{b}''(\qp)
  = \sum_{i\,<\,j} \left. \diff[2]{\Psi(j^*_{ij}, j^{p^*}_{ij}, a^{p^*}_{ij})}{q} \right|_{q=\qp}.
\end{equation}
For any choice of \(j^*_{ij}(q)\) and \(p^*_i(q)\) that satisfy \(j^*_{ij}(\qp) = \jp_{ij}\) and \(p^*_i(\qp) = \pi_i\),%
\footnote{This ensures that the inequality~\eqref{Iqbound2} is saturated at the minimum when \(q = \qp \).
Otherwise the second derivative of the bound would not necessarily be a bound on the second derivative.}
the second derivative of the rate function is given by
\begin{equation}\label{eq:densityhess}
  I''_b(\qp) = \sum_{i<j}
        \frac{1}{\phip_{ij} + \phip_{ji}}
        \brk{\diff{}{q} (j^*_{ij} - j^{p^*}_{ij})}_{q=\qp}^2.
\end{equation}
This sum can be split into the following three contributions:
\begin{equation}\label{eq:threeCont}
I''(\qp)
  \le I_{b,\CSj}''(\qp)
    + I_{b,\CNj}''(\qp)
    + I_{b,\CBj}''(\qp),
\end{equation}
where \(\CSj \) is the set of transitions between signaling states, \(\CNj \) the transitions between nonsignaling states, and \(\CBj \) the transitions from nonsignaling to signaling states.

Our choice of \(j^{*}\) must satisfy the condition \(\sum_{j} j^*_{ij} = 0\), and our choice of \(p^{*}\) must satisfy the conditions \(\sum_{i} p^*_{i} = 1\) and \(\sum_{i \in \CS} p^*_i = q\).
We also require \(j^*_{ij}(\qp) = \jp_{ij}\) and \(p^*_i(\qp) = \pi_i\).
With the benefit of hindsight, we can then choose:
\begin{equation}\label{pstar2}
\begin{aligned}
j^*_{ij}(q) &= \brk{\frac{q(1-q) + \qpm}{2\qpm}} \jp_{ij}, \\
p^*_{i}(q) &=
   \begin{cases}
      \frac{q}{\qp} \pi_{i} & i \in \CS, \\
      \frac{1-q}{\qm} \pi_{i} & i \in \CN.
   \end{cases}
\end{aligned}
\end{equation}

Defining
\begin{equation}\label{eq:EntropyDef2}
 \Sigmap_\mathcal{X}
   \equiv \sum_{\substack{i < j \\ \mathclap{(i,j) \in \mathcal{X}}}} \sigmap_{ij}
   = \sum_{\substack{i < j \\ \mathclap{(i,j) \in \mathcal{X}}}}
     \jp_{ij} \log \frac{\phip_{ij}}{\phip_{ji}}
 \end{equation}
as the steady state energy consumption rate (in units of \(\kt \)) due to transitions along edges in the sets \(\mathcal{X} = \{\CSj,\CNj,\CBj \} \), and
\begin{equation}\label{fluxdef2}
\Bindp
  \equiv \sum_{\substack{i\in \CS \\ j\in \CN}} \phip_{ij}
  = \sum_{\substack{i\in \CS \\ j\in \CN}} \phip_{ji}
\end{equation}
as the flux due to transitions from the signaling states to the nonsignaling states, we find
\begin{equation}\label{dIdqbound_1}
I_{b,\CSj}''(\qp)
  = \sum_{i<j} \frac{\phip_{ij} - \phip_{ji}}
                    {4 \qpmb^2}
  \tanh \bigg[ \frac{1}{2} \log \frac{\phip_{ij}}{\phip_{ji}} \bigg],
\end{equation}
where we made use of the identity
\( \frac{\phip_{ij} - \phip_{ji}}{\phip_{ij} + \phip_{ji}}
       = \tanh \Big[ \frac{1}{2} \log{\frac{\phip_{ij}}
                                    {\phip_{ji}}} \Big] \).
In general the inequality
\(
  \tanh \Big[ \frac{1}{2} \log{\frac{\phip_{ij}}
                                    {\phip_{ji}}} \Big]
    \leq \frac{1}{2} \log{\frac{\phip_{ij}}{\phip_{ji}}}
\)
holds, becoming an approximate equality for \objects{} near equilibrium.%
\footnote{When \(\phip_{ij} < \phip_{ji}\) the inequality is reversed, but the factor of \(\phip_{ij} - \phip_{ji}\) in \cref{dIdqbound_1} is negative in such cases.}
Applying this inequality to \cref{dIdqbound_1} and using \cref{eq:EntropyDef2}, we find,
\begin{equation}\label{dIdqbound_S}
I_{b,\CSj}''(\qp) \,
  \leq \,\frac{\Sigmap_{\CSj}}
              {8 \qpmb^2}.
\end{equation}
By the same arguments, we also find that
\begin{equation}\label{dIdqbound_NS}
I_{b,\CNj}''(\qp) \,
  \leq \, \frac{\Sigmap_{\CNj}}
               {8 \qpmb^2}.
\end{equation}
For the term contributed by transitions between the signaling and nonsignaling states, we find that
\begin{equation}\label{dIdqbound_SNS}
I_{b,\CBj}''(\qp)
  = \sum_{\substack{i\in\CS \\ j\in\CN}}
        \frac{\phip_{ij} + \phip_{ji}}
             {4 \qpmb^2}
  = \frac{\Bindp}
         {2 \qpmb^2} \, .
\end{equation}
Plugging \cref{dIdqbound_S,dIdqbound_NS,dIdqbound_SNS} into \cref{eq:threeCont}, we arrive at our final bound for the second derivative of the rate function of \(q\) evaluated at \(\qp \):
\begin{equation}\label{d2Iq}
 I''(\qp) 
 \le \frac{\Sigmap_{\CSj} + \Sigmap_{\CNj} + 4\Bindp}
          {8\qpmb^2},
\end{equation}
which implies that
\begin{equation}
  I''(\qp) 
  \le  \frac{\Sigmap_\text{tot} + 4 \Bindp}
            {8\qpmb^2} .
\end{equation}
Where \(\Sigmap_\text{tot} \equiv \Sigmap_{\CSj} + \Sigmap_{\CNj} + \Sigmap\lbj\).
Therefore, we find that the uncertainty in \(q\) is bounded by the energy consumption and flux:
\begin{equation}\label{eq:epsqbound2}
\operatorname{var}(q) \ge
  \frac{8 \qpmb^2}
       {T \left[\Sigmap_\text{tot} + 4 \Bindp \right]}.
\end{equation}
This is \cref{eq:epsqbound} in the main text.


\section{Computing the \object{} gain}\label{sec:jacobian}

\Cref{eq:stderror} from the main text shows that we need an expression for \(\diff{\qp}{c}\),
the rate of change of the \sigfrac{}, \(q\), with respect to the concentration estimate, \(\hat{c}\).
Here we present the derivation of the expression used in the main text for systems with only one nonsignaling state.
As discussed in the main text, this \object{} gain plays a role in the estimation error of the \obsq{} (\SO{}), with larger gain leading to smaller error.

Given an empirical observation of the \sigfrac{}, we can estimate the concentration by asking the question: ``For what value of \(c\) would this value of \(q\) be typical?''.
For any value of \(c\), the typical \(q\) is the one determined by the steady-state distribution: \(\qp(c) = \sum_{i\in\CS} \pi_i(c)\),
with \(\pi_i\) varying with \(c\) via the transition rates \(\MMm_{ij} \propto c\) for \(i \in \CN \), \(j \in \CS \).
Thus, the concentration estimate, \(\hat{c}(q)\) is the solution to the equation \(\qp(\hat{c}) = q\),
and therefore \(\diff{q}{\hat{c}} = \left.\diff{\qp}{c}\right\rvert_{c=\hat{c}}\).

Using the result from~\cite{cho2000markov} (see \cref{eq:diffpT}, \cref{sec:pert}) the effect of a perturbation to the rate matrix, \(\MM \), on the steady-state distribution \(\eq \) is related to the mean first-passage times, \(\fptb \), as follows:
\begin{equation}\label{eq:eqfpt}
  \diff{\pi_k}{c} = \sum_{i\neq j} \pi_i \,\diff{\MMm_{ij}}{c} \prn{\overline{T}_{ik} - \overline{T}_{jk}} \pi_k,
\end{equation}
where \(\overline{T}_{ij}\) is the mean first passage time from state \(i\) to state \(j\) for \(i \neq j\) and 0 for \(i=j\) (see \cref{sec:fpt}).

We are interested in the gradient of \(\qp = \sum_{k \in \CS} \pi_k\).
Furthermore, the only off-diagonal transition rates that depend on \(c\) are \(\MMm_{ij} \propto c\) for \(i \in \CN \) and \(j \in \CS \).
Therefore:
\begin{equation}\label{eq:jacfpt}
  \diff{\qp}{c} = \frac{1}{c} \sum_{i \in \CN} \sum_{j,k \in \CS} \pi_i \, \MMm_{ij} \prn{\overline{T}_{ik} - \overline{T}_{jk}} \pi_k.
\end{equation}
From~\cite{Yao1985} (see \cref{eq:fptbrec}, \cref{sec:fpt}), we note that \(\sum_j \MMm_{ij} \overline{T}_{jk} = \delta_{ik}/\pi_i - 1\).
Then we can write
\begin{equation}\label{eq:jacminus}
\begin{aligned}
  c \, \diff{\qp}{c}
                   &= \sum_{i \in \CN} \sum_j \sum_{k \in \CS} \pi_i \, \MMm_{ij} \prn{\overline{T}_{ik} - \overline{T}_{jk}} \pi_k
                    - \sum_{i,j \in \CN} \sum_{k \in \CS} \pi_i \, \MMm_{ij} \prn{\overline{T}_{ik} - \overline{T}_{jk}} \pi_k \\
                   &= \sum_{i \in \CN} \sum_{k \in \CS} \pi_i \pi_k
                    - \sum_{i,j \in \CN} \sum_{k \in \CS} \pi_i \, \MMm_{ij} \prn{\overline{T}_{ik} - \overline{T}_{jk}} \pi_k \\
                   &\equiv \qp \prn{1 - \qp} - A,
\end{aligned}
\end{equation}
where \(A\) is defined by this equation.

When detailed balance is satisfied, \(\pi_i \MMm_{ij}\) is symmetric in \(i,j\), whereas \(\prn{\overline{T}_{ik} - \overline{T}_{jk}}\) is antisymmetric, so \(A = 0\).
Similarly, when there is only one non-signalling state, the sum over \(i\) and \(j\) consists of one term with \(i=j\), which gives zero.
More generally, \(A=0\) if there are no transitions between nonsignaling states with nonzero rates and unbalanced fluxes.
Therefore, all detailed balanced systems \emph{and} all systems with only one non-signalling state have
\begin{equation}\label{eq:detbaljac}
  \hat{c}\diff{q}{\hat{c}} = c\diff{\qp}{c} = \qpm
  \qquad \implies \quad
  \qp(c) = \frac{1}{1 + ({K_d}/{c})}
  \qquad \implies \quad
  \hat{c}(q) = \frac{q\, K_{d}}{1-q},
\end{equation}
where \(K_{d}\) is the dissociation constant, the concentration at which \(\qp = \frac{1}{2}\).


\section{Exact error formulae for the ideal and simple observers}\label{sec:conc}

In this \namecref{sec:conc} we compute the \error{}, for the \obstraj{} and show that it saturates the Cram\'er-Rao bound (\cref{eq:trajUncertainty} in the main text).
We will then derive the expression for the \obsq{}'s \error{} that we used in numerical optimization (\cref{eq:exact,fig:fig2} in the main text).


\subsection{Error in estimating concentration: the \obstraj{}}\label{sec:mlevar}

In \cref{sec:mle} we saw that the maximum likelihood estimator could be written in terms of the empirical densities and fluxes as
\begin{equation*}
  \hat{c} = \frac{\Bindem}{\left. \Bindpem \right|_{\mathrlap{c = 1}}}.
\end{equation*}
In \cref{sec:fluxLDT} we see that the empirical densities and fluxes obey a large deviation principle.
Therefore, the concentration estimate also obeys a large deviation principle described by the contraction:
\begin{equation*}
  I_{\hat{c}}(\hat{c}) = \inf_{\pr,\boldsymbol{\phi}} I(\pr,\boldsymbol{\phi})
  \qquad \text{subject to} \quad
\begin{aligned}[t]
  &\sum_i p_i = 1, \quad
  \sum_{\mathclap{(ij) \in \CBj}} \phi_{ij}
        = \hat{c} \sum_{\mathclap{(ij) \in \CBj}}  \frac{p_i \MMm_{ij}}{c}, \quad
  \sum_{j \neq i} \phi_{ij} = \sum_{j \neq i} \phi_{ji} \;\; \forall i, \\
  &p_i \geq 0 \;\; \forall i, \quad
  \phi_{ij} \geq 0 \;\; \forall i \neq j.
\end{aligned}
\end{equation*}
We have a constrained optimization problem for each possible value of \(\hat{c}\),
so we have a Lagrangian for each value of \(\hat{c}\):
\begin{equation}\label{eq:mleLagrangian}
\begin{aligned}[t]
  \CL(\hat{c}) = I(p,\phi)
  &+ \alpha(\hat{c}) \brk{\sum_i p_i - 1}
  + \beta(\hat{c}) \smashoperator[l]{\sum_{{(ij) \in \CBj}}}
    \brk{\phi_{ij} - \frac{\hat{c}}{c}\, p_i \MMm_{ij}}
  + \sum_{i \neq j} \gamma_i(\hat{c}) \prn{\phi_{ij} - \phi_{ji}} \\
  &- \sum_i \mu_i(\hat{c}) p_i
  - \sum_{i \neq j} \nu_{ij}(\hat{c}) \phi_{ij}.
\end{aligned}
\end{equation}
where \(\alpha,\beta,\gamma_i\) are Lagrange multipliers and \(\mu_i,\nu_{ij}\) are Karush-Kuhn-Tucker multipliers, satisfying \(\mu_i \geq 0\), and \(\mu_i \partial\CL/\partial\mu_i = 0\), the allowed region being \(\partial\CL/\partial\mu_i \leq 0\), and similar for the \(\nu_{ij}\).
The Lagrange/KKT multipliers will take different values for each \(\hat{c}\) as well.

The conditions for the infimum are
\begin{equation}\label{eq:mleinf}
\begin{alignedat}{2}
  \pdiff{\CL}{p_i} &= \pdiff{I}{p_i}
    + \alpha(\hat{c}) - \frac{\beta(\hat{c}) \hat{c}}{c} e^\CN_i [\MM \onev^\CS]_i - \mu_i(\hat{c})
    &&= 0, \\
  \pdiff{\CL}{\phi_{ij}} &= \pdiff{I}{\phi_{ij}}
    + \beta(\hat{c}) e^\CN_i e^\CS_j + \gamma_i(\hat{c}) - \gamma_j(\hat{c}) - \nu_{ij}(\hat{c})
    &&= 0, \\
\end{alignedat}
\end{equation}
where \(\onev^\CS \) is a vector of ones for states in \(\CS \) and zero elsewhere,
and \(\onev^\CN\) is the reverse.%
\footnote{In general, given a vector \(\mathbf{v}\) and set of states \(\mathcal{X}\), the vector \(\mathbf{v}^\mathcal{X}\) has components
\(v^\mathcal{X}_i =
\begin{cases}
  v_i &\text{if } i \in \mathcal{X},\\
  0   &\text{otherwise}.
\end{cases}\).
}

To calculate the variance of \(\hat{c}\) for large observation time, \(T\), we only need the second derivative of the rate function at \(\hat{c} = c\).
So we Taylor expand the minimizers of \cref{eq:mleLagrangian} in \((\hat{c} - c)\) as
\begin{equation*}
  p_i = \pi_i + p'_i (\hat{c}-c) + \frac{p''_i}{2} (\hat{c}-c)^2 + \CO{(\hat{c}-c)}^3,
  \qquad
  \phi_{ij} = \phip_{ij} + \phi'_{ij} (\hat{c}-c) + \frac{\phi''_{ij}}{2} (\hat{c}-c)^2
     + \CO{(\hat{c}-c)}^2.
\end{equation*}
Because the zeroth order parts of \(p\) are nonzero, and we are only considering infinitesimal fluctuations, the inequality constraints will be loose and we can set the \(\mu_i\) to zero.
Some of the \(\phip_{ij}\) could be zero, but as we shall see below, those components do not receive any corrections and we can set the \(\nu_{ij}\) to zero as well.

The Taylor series expansion of \(I(p,\phi)\) begins at second order:
\begin{equation}\label{eq:fluxHess}
\begin{aligned}
  I(p,\phi) =  \sum_{i \neq j} \frac{\prn{\phi'_{ij} - p'_i \MMm_{ij}}^2}
                                    {2\phip_{ij}} (\hat{c}-c)^2 + \CO(\hat{c}-c)^3.
\end{aligned}
\end{equation}
Therefore, expanding \cref{eq:mleinf} to zeroth order gives
\begin{equation*}
  \alpha(c) - \beta({c}) e^\CN_i [\MM \onev^\CS]_i = 0,
  \qquad
  \beta({c}) e^\CN_i e^\CS_j + \gamma_i({c}) - \gamma_j({c}) = 0.
\end{equation*}
If we multiply the second equation by \(\MMm_{ij}\), sum over \(j \neq i\), and add the result to the first equation, we find \(\MM \bgm = \alpha \onev\).
The only solutions are \(\alpha(c) = 0\) and \(\gamma_i(c) =\)constant.
The original equations then imply that \(\beta(c) = 0\).

Then the Taylor series expansion of \cref{eq:mleLagrangian} is
\begin{equation}\label{eq:mleHess}
\begin{aligned}
  \CL =&\ \Biggl[
           \sum_{i \neq j} \frac{\prn{\phi'_{ij} - p'_i \MMm_{ij}}^2}{2\phip_{ij}}
          + \alpha' \sum_i p'_i
          + \beta' \smashoperator[l]{\sum_{{(ij)\in\CBj}}}
                \prn{\phi'_{ij} - p'_i \MMm_{ij} - \frac{\phip_{ij}}{c}}
          + \sum_{i \neq j} \gamma'_i \prn{\phi'_{ij} - \phi'_{ji}}
        \Biggr] (\hat{c}-c)^2 \\&+ \CO(\hat{c}-c)^3.
\end{aligned}
\end{equation}
If we minimize this expression with respect to \(p'_i\) and \(\phi'_{ij}\),
we find
\begin{equation*}
  \sum_{j \neq i} \frac{(p'_i \MMm_{ij} - \phi'_{ij})\MMm_{ij}}{\phip_{ij}} + \alpha' - \beta' e^\CN_i [\MM \onev^\CS]_i = 0,
  \qquad
  \frac{\phi'_{ij} - p'_i \MMm_{ij}}{\phip_{ij}} + \beta' e^\CN_i e^\CS_j + \gamma'_i - \gamma'_j = 0.
\end{equation*}
We can see that \(\alpha' = 0\) and \(\gamma_i' =\)constant with the same method used for the zeroth order parts.
This leaves us with \(\phi'_{ij} - p'_i \MMm_{ij} = \beta' e^\CN_i \phip_{ij} e^\CS_j \).
To determine \(\beta'\) we can look at the \(\beta'\) constraint in \cref{eq:mleHess}.
It shows that, to first order in \((\hat{c} - c)\), we require \(\Bind' - \Bind^{p'} =  \frac{\Bindp}{c}\) and therefore \(\beta' = \frac{1}{c}\).

We can substitute these results into \cref{eq:mleHess} to find
\begin{equation}\label{eq:mlevar}
   I''_{\hat{c}}(c) = \frac{\Bindp}{c^2}
   \qquad \implies \qquad
   \frac{\av{\delta c^2}}{c^2} = \frac{1}{\overline{N}}.
\end{equation}
This saturates the Cram\'er-Rao bound, \cref{eq:trajUncertainty} in the main text.


\subsection{Exact variance of the \sigfrac}\label{sec:varq}

First we compute the variance of the \sigfrac{} \(q\), which is the fraction of time the \object{} is bound and signalling along a single trajectory, by solving the contraction to leading order in \((q-\qp)\).
The contraction of the rate function from empirical density, \(p\), and current, \(j\), to empirical signalling density, \(q\) is
\begin{equation*}
  I_q(q) = \inf_{\pr,\jb} I(\pr,\jb)
  \qquad \text{subject to} \qquad
    \sum_i p_i = 1, \quad
    \sum_{i \in \CS} p_i = q, \quad
    \sum_j j_{ij} = 0, \quad
    p_i \geq 0.
\end{equation*}
We can find the infimum by minimizing the following Lagrangian:
\begin{equation*}
  \CL = I(p,j)
    + \alpha(q) \brk{\sum_i p_i - 1}
    + \beta(q) \brk{\sum_{i \in \CS} p_i - q}
    + \sum_{ij} \gamma_i(q) j_{ij}
    - \sum_i \mu_i(q) p_i.
\end{equation*}
where \(\alpha,\beta,\gamma_i\) are Lagrange multipliers and \(\mu_i\) are Karush-Kuhn-Tucker multipliers.
As we have an optimization problem for each possible \(q\), there will be different values of the Lagrange/KKT multipliers for each \(q\) as well.
The contraction is then determined by
\begin{equation}\label{eq:contractD}
\begin{aligned}
  \pdiff{I}{p_i} &= -\alpha(q) - \beta(q) e^\CS_i + \mu_i(q), &\qquad
  \pdiff{I}{j_{ij}} &= \gamma_j(q) - \gamma_i(q),
\end{aligned}
\end{equation}
where \(\onev^\CS \) is a vector of ones for states in \(\CS \) and zero elsewhere.

We assume that at \(q = \qp \), we have \(p_i = \pi_i\) and \(j_{ij} = \jp_{ij} = \pi_i\MMm_{ij} - \pi_j \MMm_{ji}\) (see \cref{secondDbound2}).
We also assume that the solution lies in the interior of the allowed region where \(p_i > 0\) and \(\mu_i = 0\) (for an ergodic process, all \(\pi_i\) are nonzero, and for infinitesimal \((q-\qp)\) the same will be true of \(p_i\)).
From the series expansion of \(I_q (q)\) about \(q = \qp\) and \cref{eq:densityhess} we can see that
\begin{equation}\label{eq:density2ndorder}
  I_q(q) = \frac{{(q-\qp)}^2}{2} \sum_{i<j}
        \frac{1}{\phip_{ij} + \phip_{ji}}
        \brk{\diff{}{q} (j_{ij} - j^p_{ij})}_{q=\qp}^2 + \CO{(q-\qp)}^3.
\end{equation}
Therefore, we only need the expansion of the optimal \(p,j\) to first order in \((q-\qp)\), whose coefficients we denote by \(p',j'\).%
\footnote{The choice made in \cref{sec:ldt_bnd}, \cref{pstar2} would give
\(\pr' = \frac{\eq^\CS}{\qp} - \frac{\eq^\CN}{\qm}\) and
\(\jb' = \frac{\jbp}{2} \prn{\frac{1}{\qp} - \frac{1}{\qm}}\).}
Then we can expand \cref{eq:contractD} to first order to find
\begin{equation}\label{eq:contractE}
\begin{aligned}
  \sum_{j \neq i} \frac{\MMm_{ij} (j'_{ij} - j^{p'}_{ij})}{\phip_{ij} + \phip_{ji}}  &= \alpha' + \beta' e^\CS_i, &\qquad
  \frac{j'_{ij} - j^{p'}_{ij}}{\phip_{ij} + \phip_{ji}} &= \gamma'_j - \gamma'_i.
\end{aligned}
\end{equation}
Where similarly, $\alpha'$, $\beta'$, and $\gamma'$ are the first order coefficients of the Lagrange multipliers.
The constraints on \(p\) and \(j\) (\(\pr\onev=1\), \(\pr\onev^\CS=q\), \(\jb\onev = 0\)) imply that
\begin{equation}\label{eq:contractC}
  \pr' \onev = 0,
  \qquad
  \pr' \onev^\CS = 1,
  \qquad
  \jb' \onev = 0.
\end{equation}
We can solve these equations with some tools from \cref{sec:markovprimer}.
First, we can solve for \( j'_{ij} - j^{p'}_{ij} \) in the second equation of~\eqref{eq:contractE} and insert the result into the first equation of~\eqref{eq:contractE}:
\begin{equation}\label{eq:contractEsol}
  j'_{ij} - j^{p'}_{ij} = (\phip_{ij} + \phip_{ji}) (\gamma'_j - \gamma'_i)
  \qquad \implies \qquad
  \sum_{j \neq i} \MMm_{ij} (\gamma'_j - \gamma'_i) = \alpha' + \beta' e^\CS_i.
\end{equation}
The \(\gamma'_i\) term supplies the missing \(j=i\) term from the sum.
So we can rewrite the second part of \cref{eq:contractEsol} as
\begin{equation}\label{eq:contractRec}
  \MM \bgm' = \alpha' \onev + \beta' \onev^\CS.
\end{equation}
If we premultiply by \(\eq \), we find that \(\alpha' = - \qp \beta'\).
If we premultiply by the Drazin pseudoinverse, \(\MM\dinv \) (see \cref{eq:drazin}, \cref{sec:drazin}), we find that \((\I-\onev\eq)\bgm' = \beta ' \MM\dinv \onev^\CS \).
Looking at \cref{eq:contractE}, we only care about differences of the \(\gamma'_i\), so we can shift \(\gamma'_i\) by an arbitrary constant and choose to set \(\eq\bgm' = 0\).
Then
\begin{equation}\label{eq:contractG}
  \bgm' = \beta' \MM\dinv \onev^\CS
        = \beta' (\I-\onev\eq) \fptb \Eq \onev^\CS,
\end{equation}
where \(\Pi_{ij} = \pi_i \delta_{ij}\) and \(\fptbm_{ij}\) is the mean first-passage-time from state \(i\) to \(j\) (see \cref{eq:fptdrazin}, \cref{sec:fpt}).

Now we go back to the first part of \cref{eq:contractEsol} and sum over \(j\):
\begin{align*}
  \sum_j p'_j \MMm_{ji} &=
        \sum_j (\pi_i \MMm_{ij} \gamma'_j + \pi_j \gamma'_j \MMm_{ji}), \\
\intertext{or, using the natural definition of the adjoint (see \cref{eq:adjoints}, \cref{sec:adjoint}):}
  \pr' \MM &= {(\MM \bgm')}^\dag + \bgm'^\dag \MM = \bgm'^\dag (\MM + \MM^\dag).
\end{align*}
Substituting in \cref{eq:contractG} and postmultiplying by \(\MM\dinv \):
\begin{equation}\label{eq:contractP}
\begin{aligned}
  \pr' &= \beta' \eq^\CS \MM\dinvd (\MM + \MM^\dag) \MM\dinv
        = \beta' \eq^\CS (\MM\dinv + \MM\dinvd)
        = \bgm'^\dag + \bar{\bgm}'^\dag, &
  \text{or:} \quad
  p'_i &= \pi_i (\gamma'_i + \bar{\gamma}'_i)
\end{aligned}
\end{equation}
where we defined \(\bar{\bgm}' = \beta' \MM\dinvd \onev^\CS \), \ie the quantity \(\bgm'\) but computed for the time-reversed process.
We can then determine the Lagrange multiplier \(\beta'\) using the normalization constraints, \cref{eq:contractC}:
\begin{equation}\label{eq:contractB}
\begin{aligned}
  \pr' \onev^\CS = 1
  \quad&\implies\quad&
  \eq^\CS \bgm' &= \eq^\CS \bar{\bgm}' = \frac{1}{2}\\
  \quad&\implies\quad&
  \beta' &= \frac{1}{2 \eq^\CS \MM\dinv \onev^\CS}
         = \frac{1}{2 \sum_{ij} \brk{\qmb\pi^\CS_i - \qp \pi^\CN_i} \fptbm_{ij} \pi^\CS_j}.
\end{aligned}
\end{equation}

Now we can determine \(\jb'\) using the first part of \cref{eq:contractEsol},
\begin{equation*}
  j'_{ij} 
          = (\bar{\gamma}'_i + \gamma'_j) \phip_{ij} - (\gamma'_i + \bar{\gamma}'_j) \phip_{ji},
\end{equation*}
although we do not actually need this quantity.

Instead, we note that \cref{eq:density2ndorder} depends only on \(j'_{ij} - j^{p'}_{ij}\).
By \cref{eq:contractEsol}, this can be rewritten in terms of the \(\phip_{ij}\) and \(\gamma'_i\).
We can then substitute \cref{eq:contractB,eq:contractG} into \cref{eq:density2ndorder}, to find:
\begin{equation}\label{eq:contractIpp}
\begin{aligned}
  I''_q(\qp)
    &= \sum_{i<j} (\phip_{ij} + \phip_{ji}) \prn{\gamma'_i - \gamma'_j}^2
    = \sum_{ij} \phip_{ij}\prn{\gamma'_i - \gamma'_j}^2
    \\&= -2 \sum_{ij} \phip_{ij} \gamma'_i  \gamma'_j
    = -2 \beta'^2 \eq^\CS \MM\dinv \onev^\CS
    = -\beta'.
\end{aligned}
\end{equation}
In going from the first to second line, we made use of the fact that \(\sum_i \phip_{ij} = \sum_j \phip_{ij} = 0\) when we include the diagonal terms.

The variance in the \sigfrac{} \(q\) is given by \(1/(TI''(\qp))\)~\cite{Touchette2009}, where \(T\) is the total observation time, so from \cref{eq:contractB} we have
\begin{equation}\label{eq:varq}
  \Var{q} = \frac{2 \sum_{ij} \brk{\qp \pi^\CN_i - \qmb\pi^\CS_i} \fptbm_{ij} \pi^\CS_j}{T}.
\end{equation}
We can rewrite this in terms of set-to-point mean first-passage times
\begin{equation}\label{eq:settopoint}
  \fptbm_{\mathcal{X}j} = \frac{\sum_{i \in \mathcal{X}} \pi_i \fptbm_{ij}}{\sum_{i \in \mathcal{X}} \pi_i},
\end{equation}
where each term is weighted by the conditional probability of being in state \(i\) conditional on being in the set \(\mathcal{X}\), \(\Prob\cond{x(t) = i}{x(t) \in \mathcal{X}}\) for any nonspecific time \(t\).

Then \cref{eq:varq} reads as
\begin{equation}\label{eq:varqT}
  \Var{q} = \frac{2 \qpm}{T} \sum_{j \in \CS} \prn{\fptbm_{\CN j} - \fptbm_{\CS j}} \pi_j
        = \frac{2 \qpm}{T} \sum_{j \in \CN} \prn{\fptbm_{\CS j} - \fptbm_{\CN j}} \pi_j,
\end{equation}
where we used \cref{eq:kemeny}, \cref{sec:fpt}, which implies that
\(\sum_{j\in\CN} \fptbm_{\mathcal{X}j} \pi_j + \sum_{j\in\CS} \fptbm_{\mathcal{X}j} \pi_j = \eta \),
a constant independent of the initial set \(\mathcal{X}\).

This expression simplifies dramatically when there is only one non-signalling state, so that the sum collapses to a single term
\begin{equation*}
  \Var{q} = \frac{2 \qpm^2 \fptbm_{\CS 0}}{T}.
\end{equation*}
We can interpret this result physically if we rewrite it as follows:
\begin{equation}\label{eq:genCVphys}
  \Var{q}
    = \frac{2\brk{\qpm}^2}{\overline{N} \thold / \tunbind},
\qquad \text{where:} \quad
  \thold = \frac{\qp T}{\overline{N}}, \qquad
  \tunbind = \fptbm_{\CS 0}.
\end{equation}
Here \(\thold \) is \emph{holding time}, the mean time spent in bound states during one bound interval.
Also, when there is only one nonsignaling state, the set-to-set mean first-passage time \(\fptbm_{\CS\CN} = \fptbm_{\CS 0}\) so \(\tunbind \) is the mean time until the next unbinding event given that the \object{} is currently bound.

Note that the quantity \(\tunbind \) is not the same as \(\thold \).
In the case of \(\thold\), we would condition on the \object{} having entered the bound state at the particular time, \(t_{0}\), from which we measure the holding time.
The states would then be weighted by \(\Prob\cond{x(t_0) = i}{\text{bound at}\: t_0}\), the probability that the binding transition was to state \(i\).
\begin{equation}\label{eq:unbindcond}
\begin{aligned}
  \thold &= \sum_{i\in\CS} \fptbm_{i0} \ \Prob\cond{x(t_0)=i}{\text{bound at }t_0},
    \\
  \tunbind &= \sum_{i\in\CS} \fptbm_{i0} \ \Prob\cond{x(t)=i}{x(t)\in\CS}.
\end{aligned}
\end{equation}
In \cref{eq:settopoint}, by using the steady-state distribution we effectively average over the length of time since the last binding event, whereas if we were to calculate the holding time we would condition on it being zero.
It is always the case that \( \qp T = \overline{N} \thold \), and therefore:
\begin{equation}\label{eq:varqhold}
    \Var{q}
    = \frac{2\brk{\qpm}^2}{\overline{N}} \frac{\tunbind}{\thold}.
\end{equation}

When looking at the definitions of \(\tunbind\) and \(\thold\), one might think that \(\thold \geq \tunbind\).
This is not the case, due to the difference in the probability distribution of the initial state.
We will look at an illustrative example in \cref{sec:uniring}.


\begin{figure}
  \centering
  \textbf{A.}\aligntop{\includegraphics[height=0.36\linewidth]{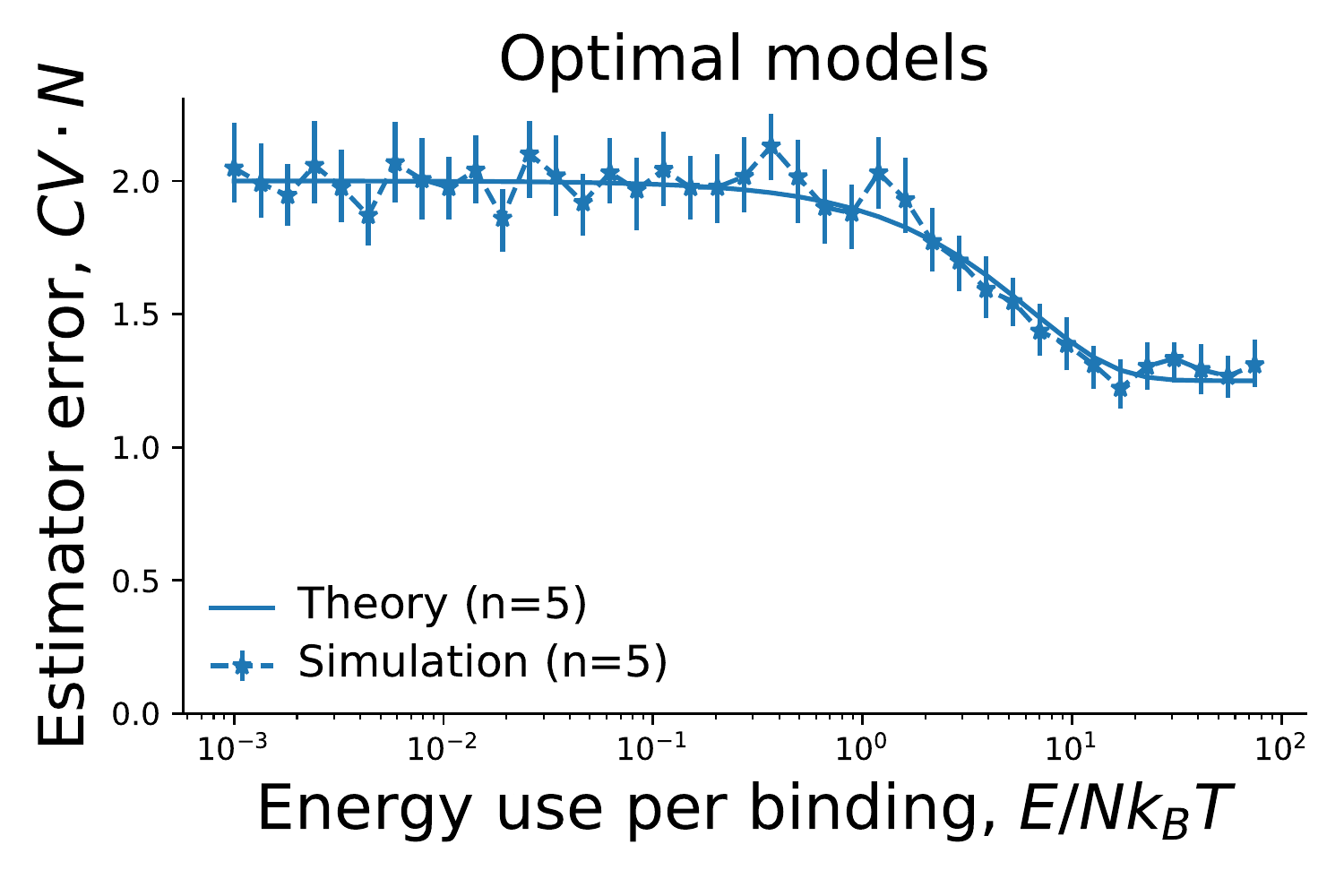}}
  \hspace{0.02\linewidth}
  \textbf{B.}\aligntop{\includegraphics[height=0.36\linewidth]{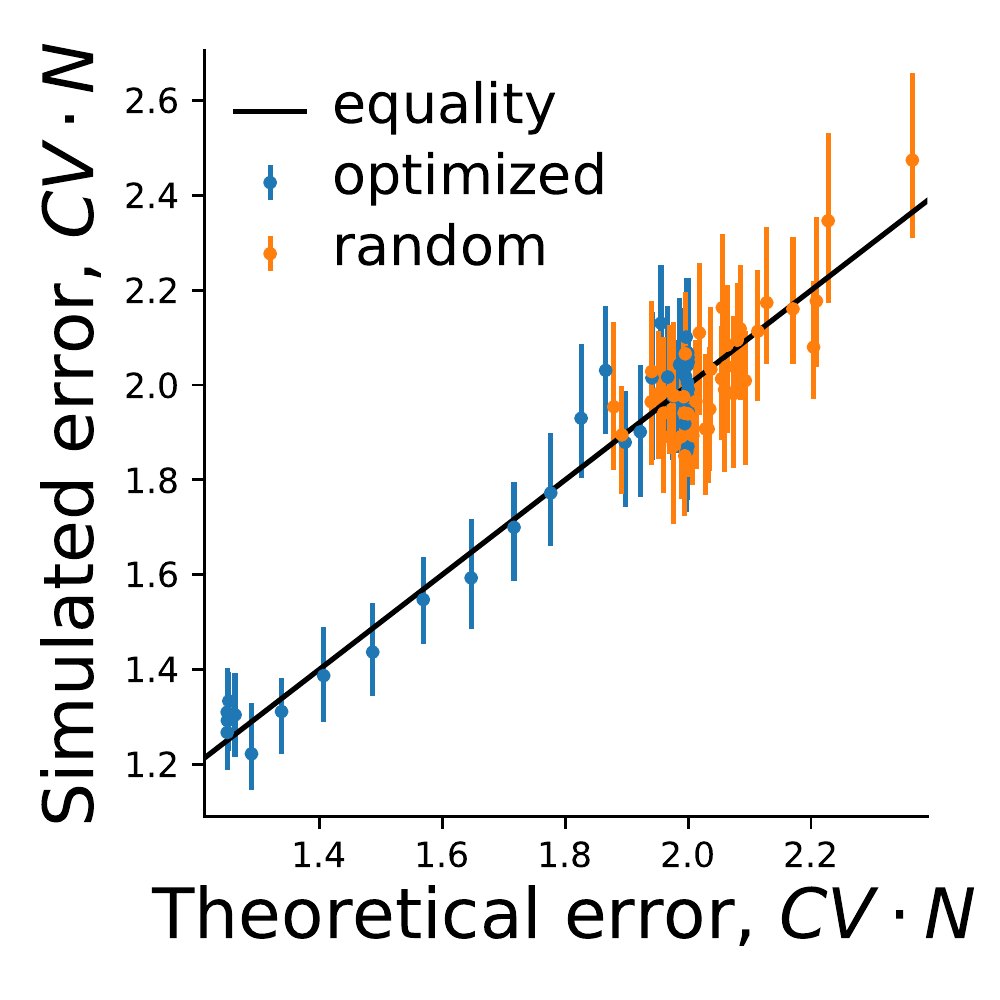}}
  \caption[Monte Carlo validation]%
  {Monte Carlo validation of \cref{eq:contractCV}.
  \textbf{A.} Comparison of the analytic expression for \error{} with Monte Carlo simulations for models with numerically optimized \(CV\) at fixed \(E\) (see \cref{fig:fig2} of the main text) and \(n=5\) states, one of which is nonsignaling.
  \textbf{B.} Comparison of the analytic expression for \error{}, \cref{eq:contractCV}, with Monte Carlo simulations for numerically optimized models from A.\ and randomly generated models.
  Simulations were performed with \(T=4000\) in units of each model's typical holding time, and 1600 repeats.
  Error bars indicate 95\% confidence intervals from 100 bootstrap resamples.
  }\label{fig:verify}
\end{figure}

\subsection{Exact error for the \obsq}\label{sec:varc}

To find the \error{} of \(\hat{c}\), we note that at the minimum of the large deviation rate function:
\begin{equation*}
  \Var{\hat{c}} = \prn{T\, \diff[2]{I}{\hat{c}}}\inv
    = \prn{T\, \diff[2]{I}{q} \brk{\diff{q}{\hat{c}}}^2}\inv
    = \frac{\Var{q}}{\brk{\diff{q}{\hat{c}}}^2}.
\end{equation*}

With only one nonsignaling state, we can use \cref{eq:detbaljac} for the jacobian between \(c\) and \(q\).
Thus:
\begin{equation}\label{eq:genCVhold}
    \frac{\Var{\hat{c}}}{c^2}
    = \frac{2}{\overline{N}} \frac{\tunbind}{\thold}.
\end{equation}
This is \cref{eq:exact} in the main text.
In the case of a two-state process (or one that is lumpable to a two-state process, see~\cite{kemeny1960finite}), \(\tunbind \) and \(\thold \) have the same distribution.
When the holding time has an exponential distribution, the time until the next unbinding is independent of the time since the last binding.
For such \objects{}, \cref{eq:genCVphys} reduces to the Berg-Purcell result~\cite{Berg1977}, \( \frac{\Var{\hat{c}}}{c^2} = \frac{2}{\overline{N}} \).

In general, we expect the \error{} to grow with the mixing time of the \object{}, as the effective number of independent observations of the \object{} scales \( \propto T / T_\text{mix} \) due to autocorrelation.
We would expect that, in most cases, a long unbinding time implies a long mixing time.

When there is more than one nonsignaling state, using \cref{eq:varq} and the jacobian from \cref{eq:jacfpt}, the long time limit of the \error{} is:
\begin{equation}\label{eq:contractCV}
  \overline{N}\, \frac{\Var{\hat{c}}}{c^2}
    = \frac{2\Bindp
             \brk{\sum_{ijk} \pi^\CN_i \pi^\CS_j \pi^\CS_k
                  \prn{\fptbm_{ik} - \fptbm_{jk}}}}
           {\brk{\sum_{ijk} \phi^{\CN\CS}_{ij} \pi^\CS_k
                 \prn{\fptbm_{ik} - \fptbm_{jk}}}^2},
\end{equation}
where \(\overline{N} = \Bindp T\) and \(\phi^{\CN\CS}_{ij}\) is \(\phip_{ij}\) for \(i \in \CN, j \in \CS \) and zero otherwise.
Given the explicit formulae for the mean first-passage-times in \cref{eq:fptbrec,eq:drazin}, the expression in \cref{eq:contractC} can immediately be computed numerically.
This is the formula that we used in numerical optimization for \cref{fig:fig2} in the main text.

We can validate \cref{eq:contractCV} with Monte Carlo simulations, as shown in \cref{fig:verify}.


\subsection{Exact first passage times and error in a uniform ring \object{}}\label{sec:uniring}

\begin{figure}[t]
   \centering
   \textbf{A.}\aligntop{\includegraphics[height=0.3\linewidth]{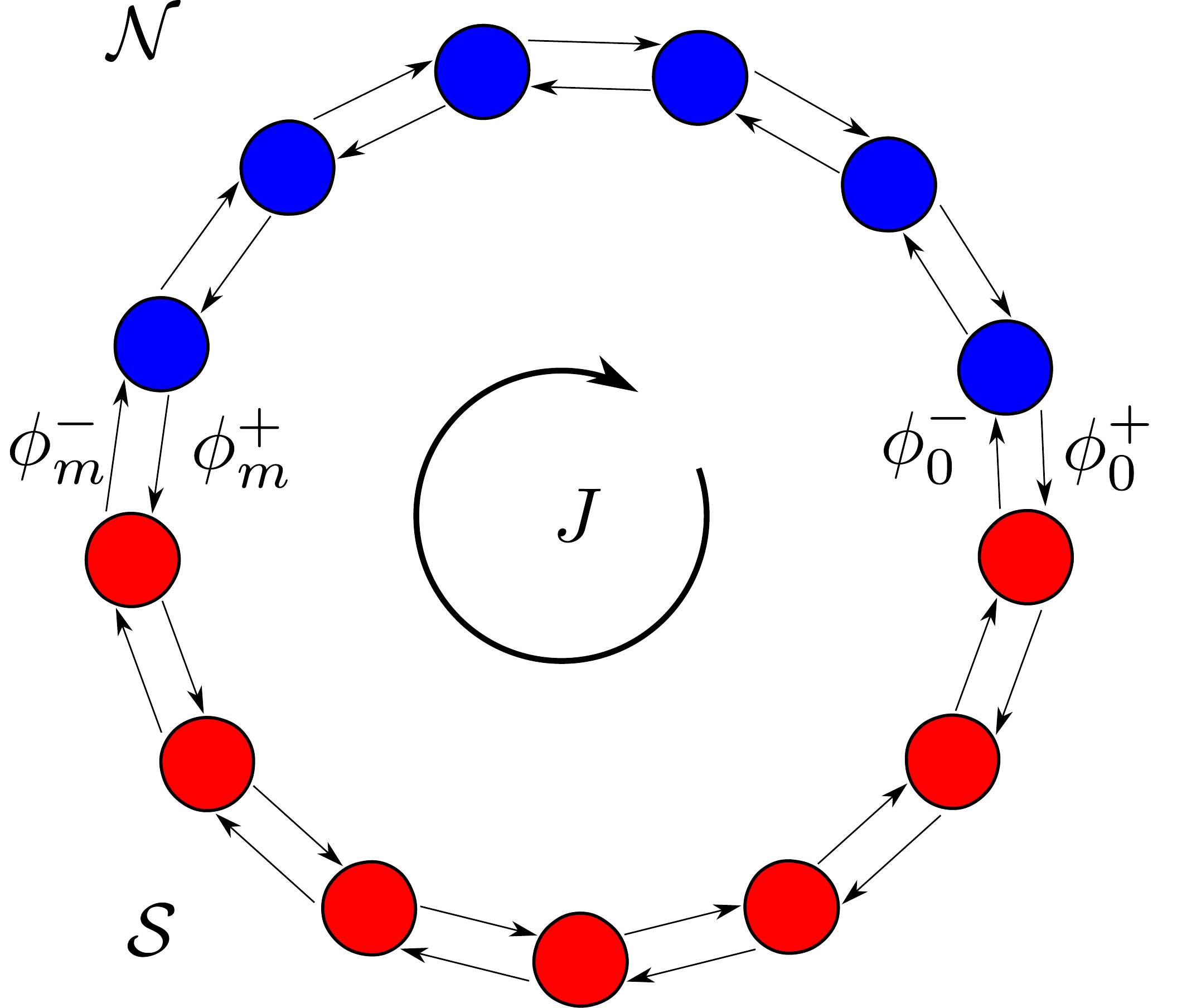}}
   \textbf{B.}\aligntop{\includegraphics[height=0.3\linewidth]{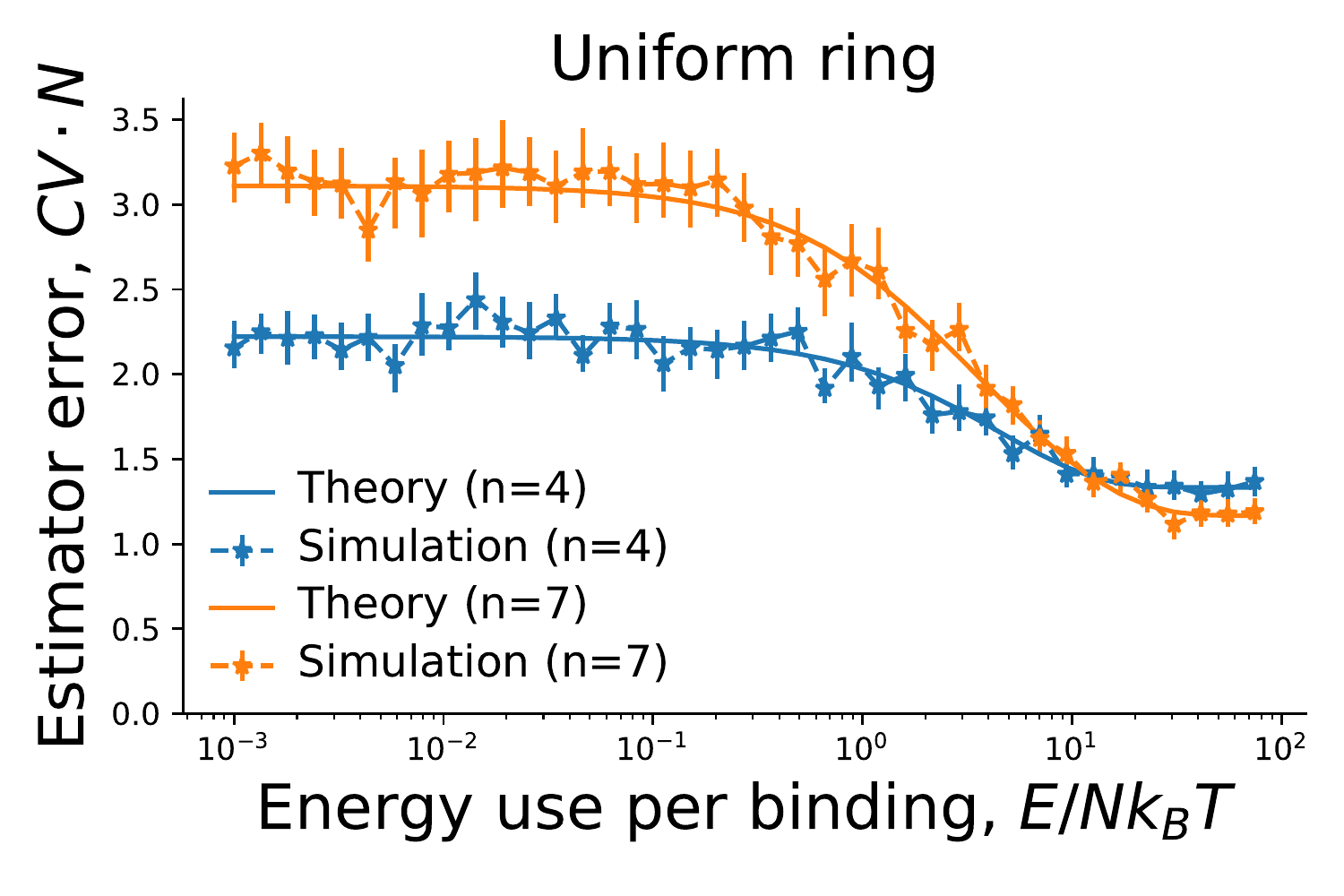}}
   \caption[Ring \objects{}]%
   {Ring \objects{}.
   \textbf{A.} An \object with a ring topology.
   \textbf{B.} Comparison of the analytic expression for \error{} of a uniform ring, \cref{eq:uniCV} with Monte Carlo simulations.
   Simulations were performed with \(T=4000\) in units of each model's typical holding time, and 1600 repeats.
   Error bars indicate 95\% confidence intervals from 100 bootstrap resamples.
   }\label{fig:ring}
\end{figure}

In this \namecref{sec:uniring} we apply \cref{eq:contractCV}, the \error{} for a general \object{}, to the case of a uniform ring \object{}.
We consider \objects{} of the type depicted in \cref{fig:ring}~A,
but with only one nonsignaling state labeled as state 0.
The transition rates are given by
\begin{equation}\label{eq:uniringdef}
  \MMm_{ij} = Q_+ \prn{\delta_{i+1,j} - \delta_{ij}}
            + Q_- \prn{\delta_{i-1j} - \delta_{ij}},
\end{equation}
where the indices are to be interpreted modulo \(n\), the total number of states.
It will be convenient to parameterize these models with the energy consumed in one full circuit of the ring
(in units of \(k_\text{B}T\)): \(\sigma = n \ln\brk{{Q_+}/{Q_-}}\).

We can determine the mean first-passage-times to the nonsignaling state using the recursion relation \cref{eq:fptbrec}
\begin{equation}\label{eq:unbindRecur}
  \MM \fptb = \Eq\inv - \onev\onev\trans,
  \quad \text{or} \quad
  Q_+(\fptbm_{i+10} - \fptbm_{i0}) + Q_- (\fptbm_{i-10} - \fptbm_{i0})
      = \begin{cases}
          -1,        & i \neq 0, \\
          \qp/\qmb,  & i = 0,
        \end{cases}
\end{equation}
whose solution is
\begin{equation*}
  \fptbm_{i0} = \frac{1}{Q_+ - Q_-}
                \brk{n \prn{\frac{1 - \e^{-i\sigma/n}}{1 - \e^{-\sigma}}} - i}.
\end{equation*}
Furthermore, the conditional probabilities in \cref{eq:unbindcond} are
\begin{equation*}
\begin{aligned}
  \Prob\cond{x(t_0)=i}{\text{bound at }t_0}
    &= \begin{cases}
      (1 + \e^{-\sigma/n})\inv, &i = 1, \\
      (1 + \e^{\sigma/n})\inv,  &i = n-1, \\
      0,  &\text{otherwise},
    \end{cases}
    \\
  \Prob\cond{x(t)=i}{x(t)\in\CS} &= \frac{1}{n-1},
  \quad i = 1, \ldots, n-1.
\end{aligned}
\end{equation*}
If we substitute these equations into \cref{eq:unbindcond}, we find
\begin{equation}\label{eq:unbinduni}
  \thold = \frac{n-1}{Q_+ + Q_-},
    \qquad
  \tunbind =
    \frac{n \prn{n \coth\brk{\frac{\sigma}{2}} - \coth\brk{\frac{\sigma}{2n}}}}
         {2(n-1)(Q_+ - Q_-)},
\end{equation}
Substituting these expressions into \cref{eq:genCVhold}, we find
\begin{equation}\label{eq:uniCV}
  \overline{N}\, \frac{\Var{\hat{c}}}{c^2}
    = \frac{n \coth\brk{\frac{\sigma}{2n}}
              \prn{n \coth\brk{\frac{\sigma}{2}} - \coth\brk{\frac{\sigma}{2n}}}}
           {{(n-1)}^2}.
\end{equation}
As \(\sigma \to \pm\infty \), this expression asymptotes to \(\frac{n}{n-1}\).
As \(\sigma \to 0\), it becomes \(2 + \frac{(n-3)(n-2)}{3(n-1)}\).

\newcommand{\En}{\mathcal{E}}
\newcommand{\Enn}{\En/2n}
\newcommand{\invxth}[1][\Enn]{\Omega(#1)}
With the same parametrization, the energy consumption per binding is given by
\begin{equation}\label{eq:uniEnt}
  \En \equiv \frac{\Sigmap}{\Bindp} = \sigma \tanh \brk{\frac{\sigma}{2n}}.
\end{equation}
We can write \cref{eq:uniCV} explicitly using the inverse function of \( x \tanh x \).
First, define a function~\( \invxth[x] \) such that
\( \invxth[x] \tanh \invxth[x] = \invxth[x \tanh x] = x \)
for all~\( x \geq 0 \).
Note that \( \coth \invxth[x] = \frac{\invxth[x]}{x} \) and
\( \coth n\invxth[x] =
\frac{(\invxth[x]+x)^n + (\invxth[x]-x)^n}
{(\invxth[x]+x)^n - (\invxth[x]-x)^n} \).
Then
\begin{equation}\label{eq:uniexcplicit}
  \overline{N}\, \frac{\Var{\hat{c}}}{c^2}
  = \frac{2n^3\, \invxth}{(n-1)^2 \,\En}
      \brk{\coth n \invxth - \frac{2 \invxth}{\En}}
      .
\end{equation}
In the limits of small and large energy consumption \cref{eq:uniexcplicit} reduces to
\begin{equation*}
  \overline{N}\, \frac{\Var{\hat{c}}}{c^2} \to
\begin{dcases}
  \frac{n(n+1)}{3(n-1)} 
      - \frac{(n+1)(n^2-4)}{90(n-1)} \, \En
      + \CO(\En^2) 
  &\text{as } \En \to 0
      ,\\
  \frac{n}{n-1} 
      + \frac{2n(n-2)}{(n-1)^2} \, \e^{-\En/n}
      + \CO\prn{\e^{-2\En/n}} 
  &\text{as } \En \to \infty
      .
\end{dcases}
\end{equation*}
In \cref{fig:ring}~B.\ we have verified \cref{eq:uniexcplicit} with Monte-Carlo simulations.

Looking at \cref{eq:unbinduni} we see that for large \(\sigma\), \(\thold > \tunbind\).
For small \(\sigma\) this is reversed, \(\thold < \tunbind\).
We can understand how this happen by looking at the mean first-passage-times, as in \cref{fig:unifpt}.
In each case, \(\tunbind\) is the arithmetic mean of the first-passage-times in \cref{fig:unifpt}~A.

When \(\sigma\) is large, the ring is approximately uni-directional.
The probability distribution of the state immediately after binding is concentrated at state \(1\).
This is where the first-passage-time \(\fptbm_{i0}\) is largest, as it must go through all of the other states before reaching \(0\).
Therefore \(\thold\) is above-average and \(\thold > \tunbind\).

When \(\sigma\) is small, the ring is symmetric between both directions.
The probability distribution of the state immediately after binding is equally concentrated in states \(1\) and \(n-1\).
This is where the first-passage-time \(\fptbm_{i0}\) is smallest, as it has a 50\% chance of going straight back to \(0\).
Therefore \(\thold\) is below-average and \(\thold < \tunbind\).

 \begin{figure}[tbhp]
  \centering
  \textbf{A.}\aligntop{\includegraphics[width=0.45\linewidth]{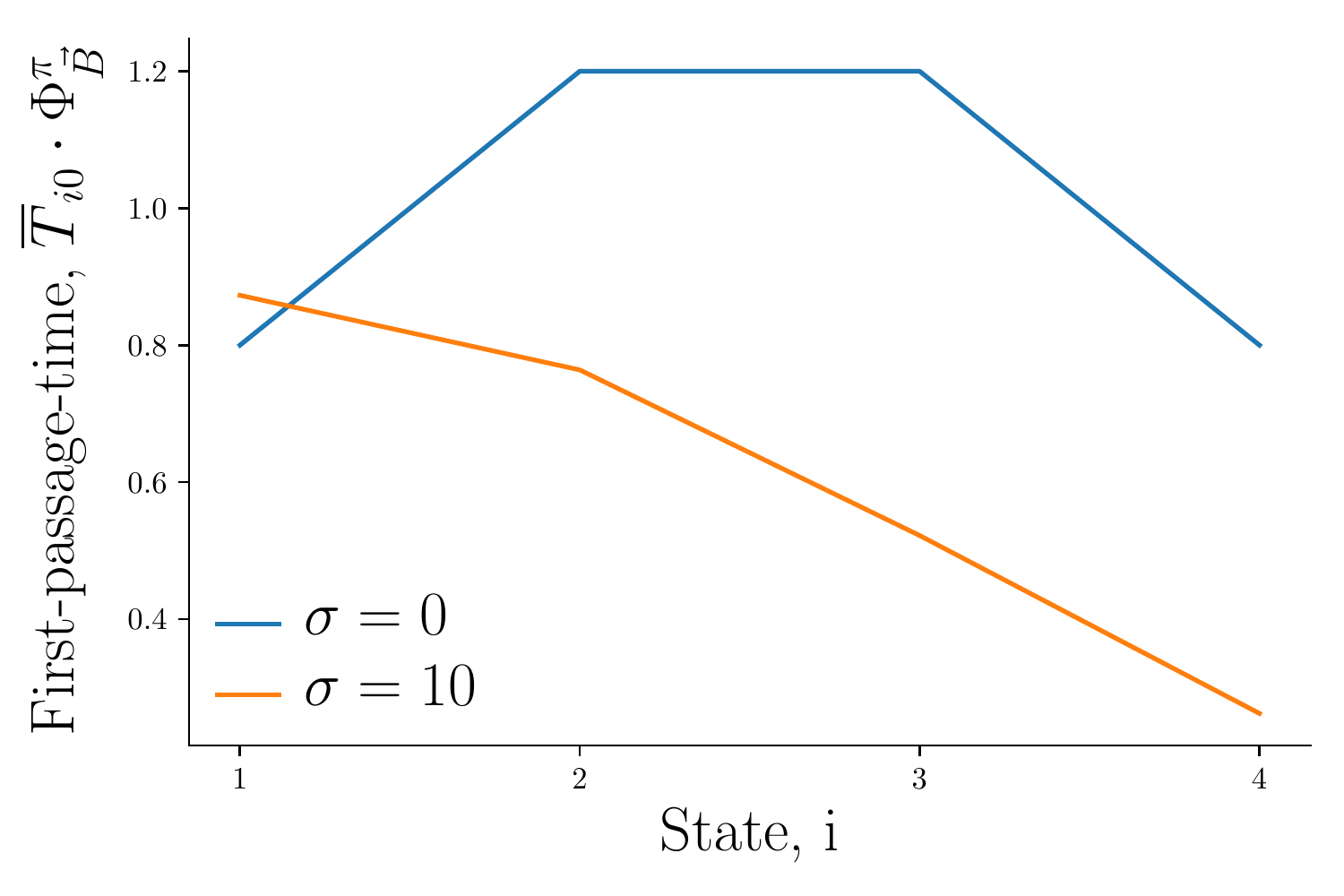}}
  \hspace{0.02\linewidth}
  \textbf{B.}\aligntop{\includegraphics[width=0.45\linewidth]{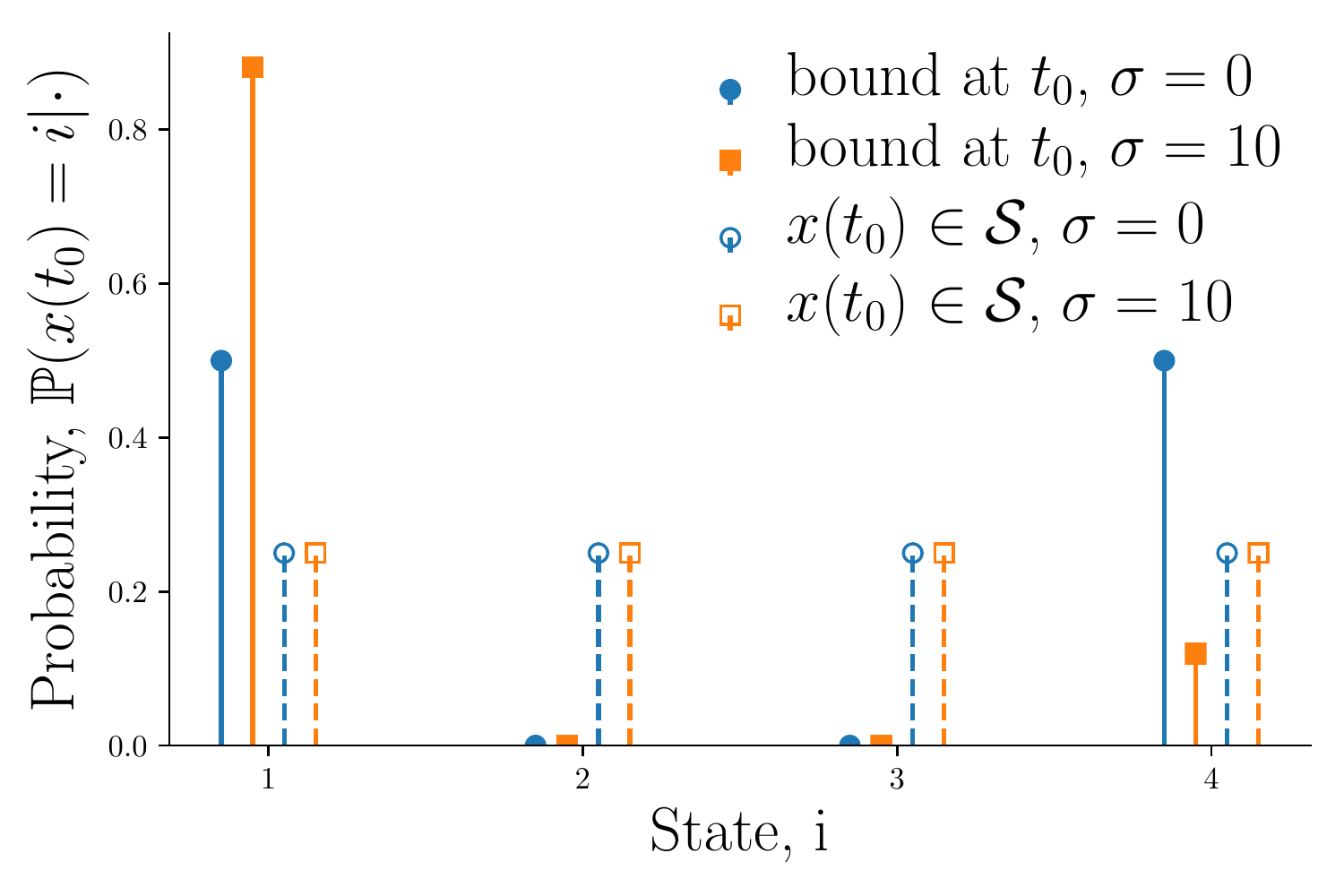}}
\caption[First-passage-times for uniform ring \objects.]{%
  First-passage-times for uniform ring \objects{} with \(n=5\).
  \textbf{A.} Mean first-passage-time from each state to state 0, the only state in \(\CN\).
  \textbf{B.} Initial probability distribution (solid) immediately after the ligand has bound, (dashed) at a generic time after binding.
  \label{fig:unifpt}}
\end{figure}


\section{Numerical Methods}\label{sec:numerics}

Here we explain in detail how we obtained the results of \Cref{fig:fig2} in the main paper, which contains numerical results falling into two categories:  results for optimized networks, and results of directly simulating randomly generated networks.

\begin{figure}[h!]
  \centering
  \aligntop{\includegraphics[width=0.85\linewidth]{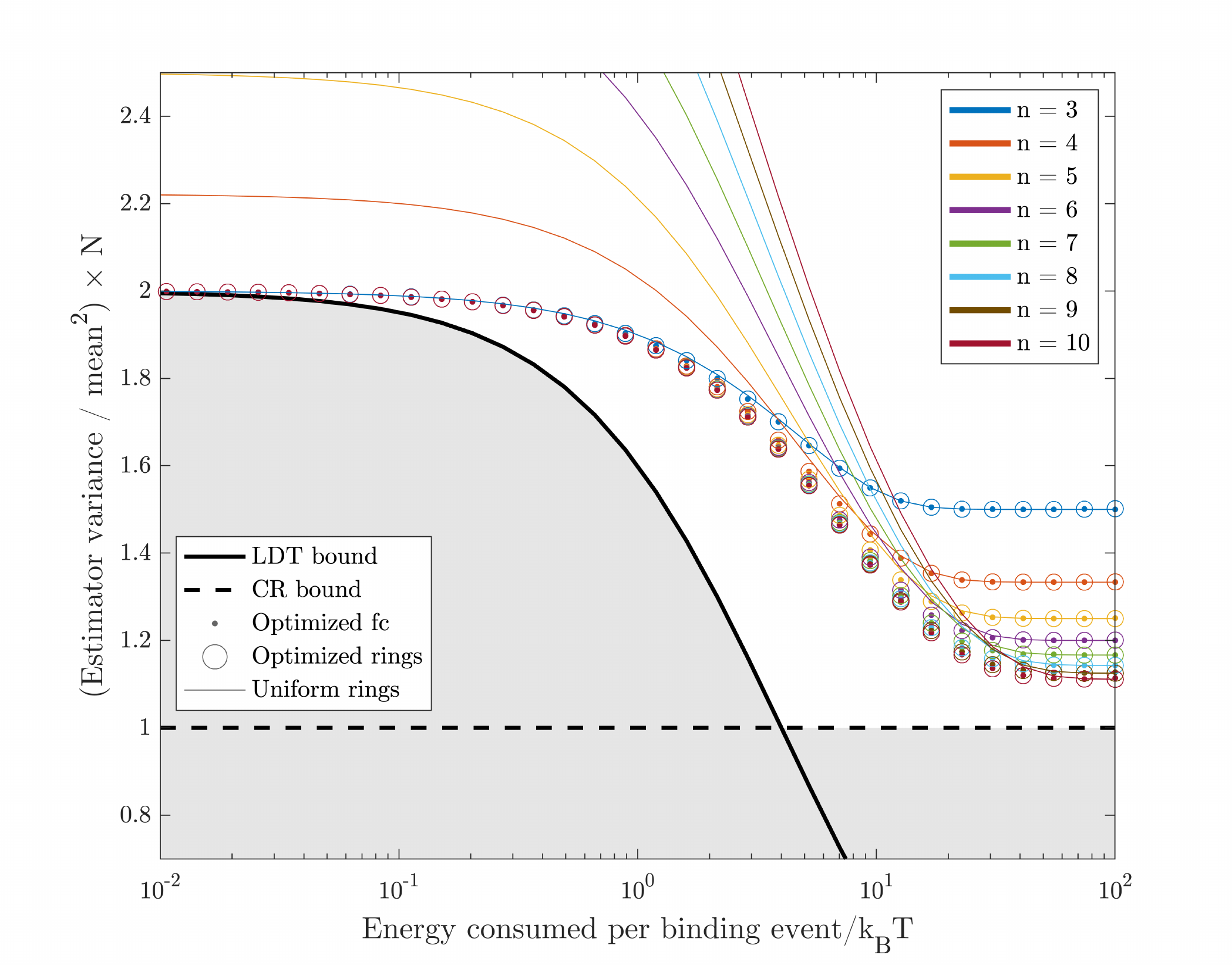}}
 \caption[Numerically optimized \objects{}]
  {\label{fig:fixedn} Numerical optimized and uniform ring networks.   
  Small solid points represent the minimal error achieved by $n$-state fully connected (fc) \objects{}, obtained by numerically minimizing~\cref{eq:optimization} with respect to all transition rates, subject to an energy constraint.
  Open circles show the same minimal error is achieved by similarly optimizing $n$-state receptors restricted to ring topologies.
  Thin solid lines represent analytic uniform ring solutions for varying $n$, \cref{eq:uniCV}.}
\end{figure}

\subsection{Numerical optimization of \objects{}}\label{sec:optNetworks}

In order to validate our theoretical bounds (\cref{eq:trajUncertainty} and \cref{eq:finalResult} in the main paper), we numerically generate networks that minimize the the exact formula (\ref{eq:contractCV}) subject to an energy consumption constraint.
The optimization problem is then:
\begin{equation}\label{eq:optimization}
\begin{split}
&\text{minimize }  \overline{N}\, \frac{\Var{\hat{c}}}{c^2}
    = \frac{2\Bindp
             \brk{\sum_{ijk} \pi^\CN_i \pi^\CS_j \pi^\CS_k
                  \prn{\fptbm_{ik} - \fptbm_{jk}}}}
           {\brk{\sum_{ijk} \phi^{\CN\CS}_{ij} \pi^\CS_k
                 \prn{\fptbm_{ik} - \fptbm_{jk}}}^2}, \\
                  &\text{subject to } \frac{\Sigmap}{\Bindp} = \text{constant}
                \end{split}
\end{equation}
\Objects{} of a given number of states and division between signaling and nonsignaling states were optimized using the MATLAB built-in nonlinear optimizing function \texttt{fmincon}~\cite{MATLAB:2017b}.
The interior-point algorithm was used to minimize the objective function in (\ref{eq:optimization}) starting from randomly initialized transition rates in a complete graph.
At each energy consumption constraint, the data point presented in \cref{fig:fig2} in the main text represents the network found giving the minimum error out of 200 optimizations with different random initializations.

\begin{figure}[b!]
  \centering
  \aligntop{\includegraphics[width=0.75\linewidth]{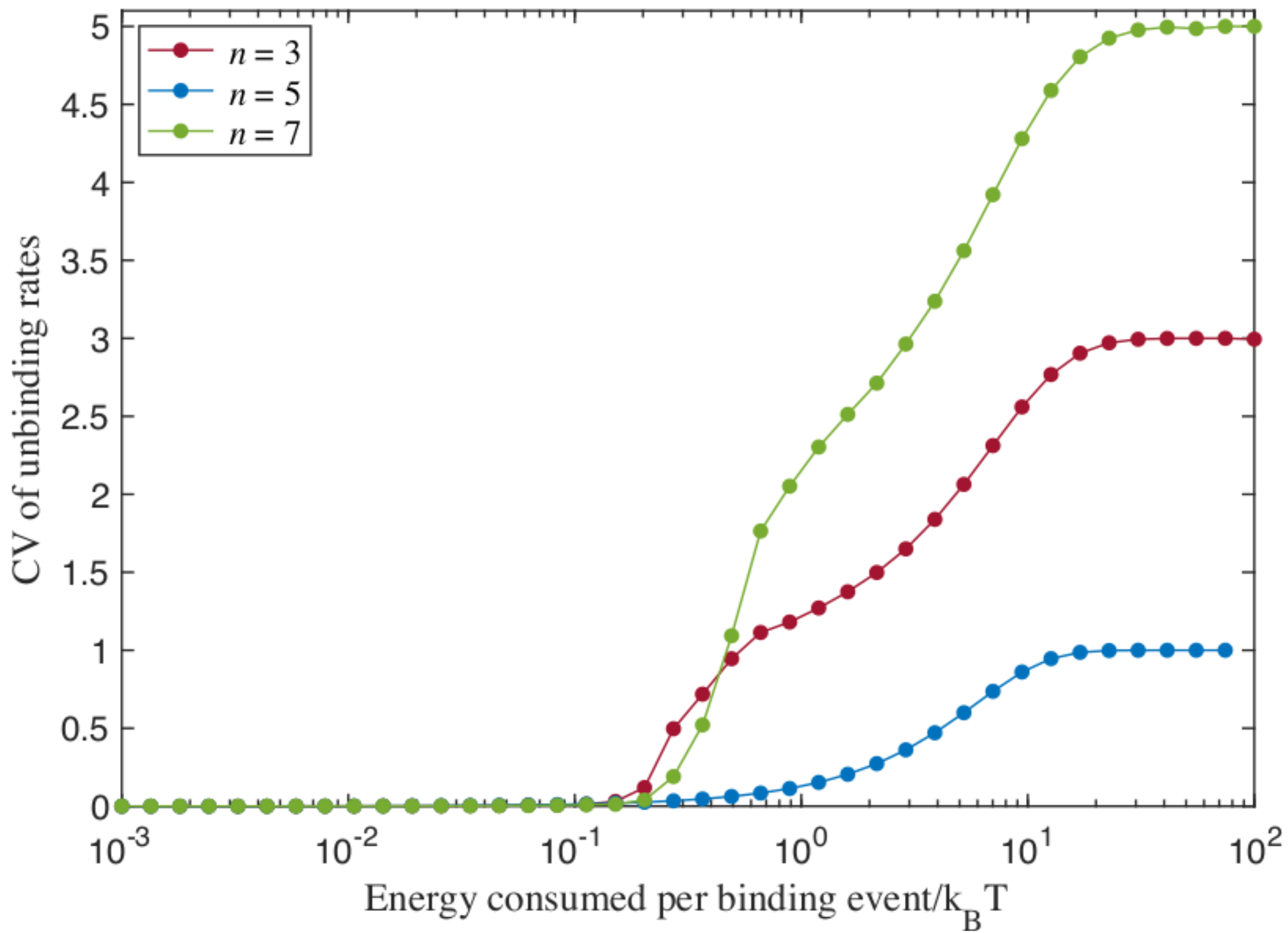}}
 \caption[Numerically optimized \objects{}]
  {\label{fig:lump} CV of the unbinding rates for $n$-state optimized networks as a function of energy dissipation per binding event.
  Networks that are lumpable to a two-state system would have CV = 0.}
\end{figure}

\subsubsection{Lumpability of optimized networks}\label{sec:lump}

Lumpability~\cite{kemeny1960finite} is a property of certain continuous-time Markov chains which indicates that the size of the state space can be reduced by `lumping' together states according to a certain partitioning.
A lumped state, which represents some subset of original states, will obey the same exponentially distributed holding time as the original subset.
Let a continuous-time Markov chain with states $\mathcal{V}$ have a partitioning of states into $n$ disjoint subsets $\{\mathcal{A}_1, \mathcal{A}_2, \ldots, \mathcal{A}_n\}$.
The Markov chain is \emph{lumpable} with respect to the partitioning if the transition rates $Q_{ij}$ from state $i$ to state $j$ obey the following:

\begin{equation}
\sum_{j\in \mathcal{A}_\ell} Q_{ij}  = \sum_{j\in \mathcal{A}_\ell} Q_{kj},  \:\: \forall i,k \in \mathcal{A}_m
\end{equation}

for any pairs of subsets in the partitioning (values of $\ell$ and $m$).
Under this condition, due to the memoryless nature of the exponential distribution, the probability of transition out of a subset $\mathcal{A}_m$ is independent of the microscopic details of which state in $\mathcal{A}_{m}$ the system occupies.
The lumped chain formed by the partitioning is then also a Markov chain with transition rate between $\mathcal{A}_m$ and $\mathcal{A}_\ell$ given by $\sum_{j\in \mathcal{A}_\ell} Q_{ij}$ for $i \in \mathcal{A}_m$.

It is potentially interesting to consider whether the optimal networks for concentration estimation are lumpable to two state processes, along the partitioning into nonsignaling and signaling states.
To measure the lumpability, we calculate the variance over the mean squared (uncertainty or CV) of the unbinding rates $Q_{i0}$, where $0$ indicates the one nonsignaling state.
If the process is perfectly lumpable, this uncertainty will be 0.
For an $n$-state uni-directional cyclic process with uniform transition rates, the CV will be $n-2$.
\Cref{fig:lump} shows the CV of unbinding rates for $n$-state Markov processes found to be optimal for concentration estimation, as a function of energy dissipation per binding event.
All processes are approximately lumpable to a two-state system until a critical dissipation level, where they separate, eventually saturating at $n-2$ as the optimal processes are all uniform rings.

\subsection{Numerical optimization of \objects{} with \texorpdfstring{\(>1\)}{>1} nonsignaling state}\label{sec:varyN}

The optimization problem described in \cref{eq:optimization} applies to \objects{} with arbitrary partitioning between signaling and nonsignaling states.
However, we find that for all models examined the best performing partition for $n$-state \objects{} are those with 1 nonsignaling state.
This is shown in \cref{fig:varyN} for all partitionings of $n = 5$ node \objects{}.

\begin{figure}[h]
  \centering
  \aligntop{\includegraphics[width=0.8\linewidth]{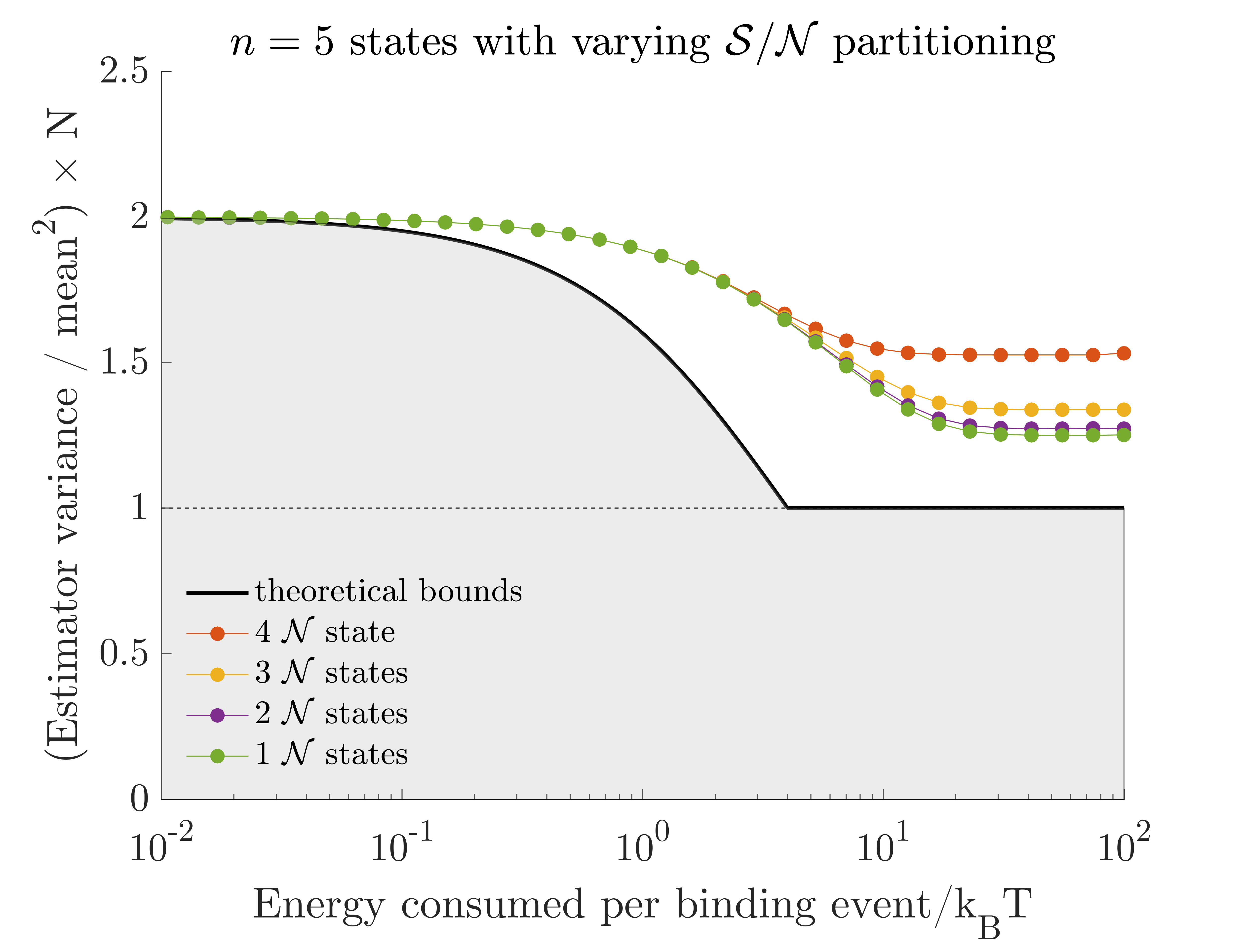}}
 \caption[Numerically optimized \objects{}]
  {\label{fig:varyN} Numerically optimized \error{} for $5$-state networks as a function of energy dissipation per binding event.
  Partitioning of states into signaling ($\mathcal{S}$) and nonsignaling ($\mathcal{N}$) is varied, with the minimal curve achieved by the single $\mathcal{N}$-state network.}
\end{figure}


\section{Extension to nonlinear dependence}\label{sec:nonlinear}

The assumption of binding transition rates that are linear in concentration may be expected biophysically for processes involving a single ligand binding.
For processes that require binding at multiple sites we can expect power law dependences.
For example the SNARE complex in neuronal synapses requires 5 calcium ions to bind for vesicle docking, the rate of which depends on calcium concentration to a power between 4 and 5~\cite{bollmann2000calcium,schneggenburger2000intracellular}.
If all the binding transitions follow the same power law, that would scale \cref{eq:trajUncertainty} of the main text and \cref{eq:detbaljac} by a constant:
\begin{equation}\label{eq:power}
  \MMm_{ij} \propto c^{k}, \; ij \in \CBj
  \quad \implies \;
  J_{c} = J_{c}^{0} + \frac{k^2 \Bindp T}{c^2},
  \quad
  c\diff{\qp}{c} = k\, \qpm,
  \quad
  \epsilon_c^2(k) = \frac{\epsilon_c^2(1)}{k^2}.
\end{equation}
Equivalently, we can say that uncertainty of the quantity \(c^{k}\) obeys all of the equations and inequalities in this work.
If the binding transitions are related to the concentration through an arbitrary function, \(f(c)\), we will find using \cref{eq:FisherGeneral} a different dependence of the Fisher information on the concentration.
The resulting uncertainty may depend then on the value of \(c\) being estimated.
However, all of our results would apply to the uncertainty in \(f(c)\).

If all of the binding transitions are potentially different functions of \(c\), then we will find unequal weights, \(W_{ij}\), in the Fisher information associated with those transitions (in \cref{eq:power} \(W_{ij} = k\)):
\begin{equation}\label{eq:nlfisher}
  W_{ij}(c) \equiv \frac{c\,\MMm_{ij}'(c)}{\MMm_{ij}(c)}
  \qquad\implies\qquad
  J_{c} = J_{c}^{0} + \frac{T}{c^2} \sum_{ij \in \CBj} W_{ij}^2\,
  \pi_{i} Q_{ij}.
\end{equation}
In this case, observations of the identity of the binding transition used can be informative of the concentration signal.

Furthermore, in this case the \SO{} would be affected only through the gain, \( c\diff{\qp}{c} \):
\begin{equation}\label{eq:nljac}
\begin{aligned}
  c\, \diff{\qp}{c} = \sum_{i \in \CN} \sum_{j,k \in \CS} W_{ij}\,
  \pi_i \, \MMm_{ij} \prn{\fptbm_{ik} - \fptbm_{jk}} \pi_k .
\end{aligned}
\end{equation}
This would modify the denominator of our expression for the true uncertainty, \cref{eq:contractCV}, as well as our lower bound, \cref{eq:finalResult} of the main text.
Cases where all of the binding transitions are potentially different functions are particularly important when generalizing our results to sensing quantities other than concentration.

\newpage
\begin{appendices}\label{sec:appendix}

\section{Primer on continuous-time Markov processes}\label{sec:markovprimer}

In this \namecref{sec:markovprimer} we provide a quick summary of those aspects of the theory of Markov processes in continuous time that are used in this supplement.

In the following sections we describe the \hyperref[sec:mastereqn]{transition rate matrix}, its \hyperref[sec:drazin]{Drazin pseudoinverse}, its relation to \hyperref[sec:fpt]{mean first-passage times}, the relation between first passage times and \hyperref[sec:pert]{perturbations of the steady-state distribution}, and the natural definition of the \hyperref[sec:adjoint]{inner product and adjoint} for vectors on the Markov chain state-space.


\subsection{Master equation and the transition rate matrix}\label{sec:mastereqn}

We consider a Markov process on a discrete set of \(n\) states indexed by \(i = 1,\ldots,n\).
We describe this process by a set of \emph{transition rates} \(\MMm_{ij}\) denoting the probability per unit time that the system jumps to state \(j\) given that it is currently in state \(i\).
The probability that the system is in state \(i\) at time \(t\), \(p_i(t)\), evolves according to the \emph{master equation}:
\begin{equation}\label{eq:master}
  \diff{p_i}{t} = \sum_{j \neq i} \left[ p_j(t) \MMm_{ji} - p_i(t) \MMm_{ij} \right].
\end{equation}
The master equation can be written in matrix form if we let \(p_i\) be the components of a row vector \(\pr(t)\) and we define the diagonal elements of the \emph{transition rate matrix} as
\begin{equation}\label{eq:mastermatrix}
  \MMm_{ii} = - \sum_{j \neq i} \MMm_{ij} \equiv - \lambda_i,
  \qquad
  \diff{\pr(t)}{t} = \pr(t) \MM.
\end{equation}
The quantity \(\lambda_i\) is the probability per unit time that the system jumps to any other state given that it is currently in state \(i\).
The holding time, or amount of time spent in any individual visit to state \(i\), follows an exponential distribution with mean \(1/\lambda_i\).
The probability that the next state visited by the Markov process is state \(j\), given that it is currently in state \(i\), is given by \(\Pbm_{ij} = \frac{\MMm_{ij}}{\lambda_i}\).

The definition of the diagonal elements in \cref{eq:mastermatrix} imply that the sum of matrix elements across any row of \(\MM \) is zero.
If we define \(\onev \) to be a column vector of ones, we can express the row sums as \(\MM\onev=\mathbf{0}\).

The \emph{steady-state distribution} \(\eq \) is the solution of \(\diff{\pr}{t} = \mathbf{0}\), and thus obeys:
\begin{equation}\label{eq:steadystate}
  \eq \MM = \mathbf{0},
  \qquad
  \eq \onev = 1.
\end{equation}
For an ergodic process \(\eq \) is uniquely defined by \cref{eq:steadystate} and is strictly positive in every state.

For future use, it will be helpful to define diagonal matrices, \(\Lb \) and \(\Eq \), with
\begin{equation}\label{eq:diags}
  \Lambda_{ij} = \lambda_i \delta_{ij},
  \qquad
  \Pi_{ij} = \pi_i \delta_{ij}.
\end{equation}
Then the matrix of next-state probabilities, \(\Pbm_{ij}\), can be written as:
\begin{equation}\label{eq:nextstate}
  \Pbm_{ij} = \Prob\cond{x_{r+1}=j}{x_r=i}
    = \begin{cases}
        0,                                    & i = j, \\
        \frac{\MMm_{ij}}{\sum_{k} \MMm_{ik}}, & \text{otherwise,}
      \end{cases}
  \qquad \text{or} \qquad
  \Pb = \I +\Lb\inv \MM.
\end{equation}
%


\subsection{Drazin pseudoinverse}\label{sec:drazin}

The transition rate matrix \(\MM \) of an ergodic Markov process has a one dimensional null-space (because \(\eq\MM \) and \(\MM\onev \) are both zero).
Therefore the rate matrix is not invertible.
However there are several ways of defining a pseudoinverse~\cite{hunter2000survey}.
The most useful one for our purposes is the \emph{Drazin pseudoinverse} of \(\MM \), defined by
\begin{equation}\label{eq:drazin}
  \MM\dinv = \tau \onev \eq - \prn{\frac{\onev \eq}{\tau} - \MM}\inv,
  \qquad
  \MM\dinv \MM = \MM \MM\dinv = \I - \onev \eq,
\end{equation}
where \(\tau \) is an arbitrary timescale.
The Drazin pseudoinverse, \(\MM\dinv \), has the same left and right eigenvectors and null spaces as \(\MM \), but with nonzero eigenvalues inverted.
In particular, \(\MM\dinv \onev = \mathbf{0}\) and \(\eq \MM\dinv = \mathbf{0}\).


\subsection{Mean first-passage times}\label{sec:fpt}

We define the \emph{mean first-passage time}, \(T_{ij}\), as the mean time it takes the process to reach state \(j\) for the first time, starting from state \(i\).
The diagonal elements, \(T_{ii}\), are defined to be the mean time it takes the process to leave and then return to state \(i\).
It will be convenient to additively decompose the mean first-passage-time matrix into its diagonal and off-diagonal parts: \(\fpt = \fpt\dg + \fptb \).

To compute the mean first passage times, consider the first time the process leaves state \(i\).
On average, this will take time \(\lambda_{i}\inv \).
With probability \(\Pbm_{ij}\), it will go directly to \(j\), so the conditional mean time would be \(\lambda_{i}\inv \).
On the other hand, if it goes to some other state, \(k\), with probability \(\Pbm_{ik}\), the conditional mean time would be \(\lambda_{i}\inv + T_{kj}\).
Combining these, we get the recursion relation
\begin{equation}\label{eq:fptrecursion}
\begin{aligned}
  T_{ij} &= \sum_{k\neq j} \Pbm_{ik} (\lambda_{i}\inv + T_{kj})
            + \Pbm_{ij} \lambda_{i}\inv
  \\     &= \sum_{k\neq j} \Pbm_{ik}T_{kj} + \sum_{k} \Pbm_{ik}\lambda_{i}\inv
  \\     &= \sum_{k} \Pbm_{ik} \fptbm_{kj} + \lambda_{i}\inv,
  & \text{or} \quad
    \fpt &= \Pb \fptb + \Lb\inv \onev\onev\trans,
\end{aligned}
\end{equation}
where \(\onev\onev\trans \) is the matrix of all ones.
Remembering \cref{eq:nextstate} (that \(\Pb = \I +\Lb\inv \MM \)), we can write \cref{eq:fptrecursion} as
\begin{equation}\label{eq:fptrecdg}
  \Lb \fpt\dg = \MM \fptb + \onev\onev\trans.
\end{equation}

The recurrence times are given by
\begin{equation}\label{eq:recurtime}
  T\dg_{ii} = \frac{1}{\pi_i\lambda_i}.
\end{equation}
This can be proved by pre-multiplying \cref{eq:fptrecdg} by \(\eq \) and employing \(\eq\MM =  0\) and \(\eq \onev = 1\)

We can substitute \cref{eq:recurtime} into \cref{eq:fptrecdg} to get a recursion relation for the off diagonal part (see~\cite{Yao1985}):
\begin{equation}\label{eq:fptbrec}
  \MM \fptb = \Eq\inv - \onev\onev\trans.
\end{equation}
Because we require that \(\fptb \) is zero on the diagonal, and the only null vector of \(\MM \) is nonzero everywhere, \cref{eq:fptbrec} has a unique solution given by
\begin{equation}\label{eq:fptdrazin}
  \fptbm_{ij} = \frac{\MMm\dinv_{ij}-\MMm\dinv_{jj}}{\pi_j}.
\end{equation}
This equation can also be written as \(\MM\dinv = (\I - \onev\eq) \fptb \Eq \).

This expression for \(\fptb \) leads to \emph{Kemeney's constant}  \(\eta \) given by~\cite{kemeny1960finite}:
\begin{equation}\label{eq:kemeny}
  \eta = \sum_j \fptbm_{ij} \pi_j.
\end{equation}
That \(\eta \) is indeed a constant reflects the remarkable fact that the quantity \(\sum_j \fptbm_{ij} \pi_j\) is actually \emph{independent} of the initial state \(i\).
If we substitute \cref{eq:fptdrazin} in, we find that \(\eta = -\Tr\MM\dinv \).


\subsection{Perturbations of the steady state distribution}\label{sec:pert}

Suppose the transition rate matrix \(\MM \) is a function of some parameter \(\alpha \).
By \cref{eq:steadystate}, \(\eq \) will also be a function of \(\alpha \).
If \(\alpha \) is changed by a small amount, \(\eq \) will also change.
This change can be expressed in terms of first-passage-times, as shown in~\cite{cho2000markov}
\begin{equation}\label{eq:diffpT}
\begin{aligned}
  \diff{\pi_k}{\alpha}
   &= \sum_{i\neq j}  \pi_i \diff{\MMm_{ij}}{\alpha}
                                   \prn{\fptbm_{ik} - \fptbm_{jk}} \pi_k.
\end{aligned}
\end{equation}
This result can be proved by expanding \(\diff{}{\alpha} (\eq\MM) = 0\), postmultiplying by \(\MM\dinv \) and using the identities from \cref{sec:drazin,sec:fpt}.
Note that the summand vanishes for \(i=j\), so we could drop the restriction \(i\neq j\) from the range of the sum.


\subsection{Inner products and adjoints}\label{sec:adjoint}

It is useful to define a natural inner product and associated norm on the space of functions over Markov chain states.
To  motivate this, it is useful to first consider inner products of functions over infinite or continuous state spaces.
The constant function, corresponding in the discrete case to the vector of all ones, \(\onev \), plays such a fundamental role that it is important that its norm, \(\nrm{\onev}\), be finite.
In order to achieve any such finite norm for a constant function over an infinite space, one requires a distribution against which to integrate the function, or compute inner products.

Returning from continuous state-spaces to discrete state-spaces, functions over continuous space correspond to column vectors over discrete states and distributions over continuous spaces correspond to row vectors over discrete states.
In the context of Markov processes, a natural such distribution is the
steady-state distribution corresponding to the row vector \(\eq \).
We thus define the  \emph{inner product} \(\av{\uv, \vv}\) over a pair of column vectors \(\uv \) and \(\vv \) using the natural distribution \(\eq \):
\begin{equation}\label{eq:inner}
  \av{\uv, \vv} = \sum_i \pi_i u^\ast_i v_i = (\uv^\ast)\trans \Eq \vv,
\end{equation}
where \(u^\ast_i\) is the complex conjugate of \(u_i\) and \(\Eq = \diag(\eq)\).
This inner product defines an associated norm \(\nrm{\vv} = \sqrt{\av{\vv,\vv}}\) and under this norm the constant function \(\onev \) has norm \(\nrm{\onev} = 1\).

The \emph{adjoints} \({(\cdot)}^\dag \) of column vectors \(\mathbf{u}\), row vectors \(\rowv \) and operators \(\M \) are defined by
\begin{equation}\label{eq:adjoints}
\begin{aligned}
  \uv^\dag \vv &= \av{\uv, \vv} \; \; \forall \vv, &\quad
  \av{\rowv^\dag, \uv} &= \rowv \uv \; \; \forall \uv, &\quad
  \av{\M^\dag \uv, \vv} &= \av{\uv, \M \vv} \; \; \forall \uv,\vv \\
  {(\uv^\dag)}_i &= \pi_i u^\ast_i, &
  {(\rowv^\dag)}_i &= \frac{\xi^\ast_i}{\pi_i}, &
  {(\M^\dag)}_{ij} &= \frac{\pi_j M^\ast_{ji}}{\pi_i}.
\end{aligned}
\end{equation}
Note that \(\onev^\dag = \eq, \eq^\dag = \onev \) and the adjoint of a transition matrix is its time-reversal, which we next explain.
In discrete time, Bayes rule states that the probability of the previous state given the current state is
\begin{equation*}
  \Prob\cond{x_r}{x_{r+1}} = \frac{\Prob\cond{x_{r+1}}{x_r} \Prob\event{x_r}}
                                {\Prob\event{x_{r+1}}}.
\end{equation*}
For a system in its steady state \(\Prob\event{x_{r+1}=i} = \Prob\event{x_{r}=i} = \pi_i\).
If the transition probabilities are given by \(\Prob\cond{x_{r+1}=j}{x_r=i} = \Mdm_{ij}\), the time-reversed process obtained via Bayes rule then has transition probabilities \(\Mdm^\dag_{ji}\), defined in \cref{eq:adjoints}.

The equivalent statement in continuous time follows from \({\exp(\MM t)}^\dag = \exp(\MM^\dag t)\).
This can be seen by noticing that this adjoint obeys the usual product rule \({(\mathbf{A}\mathbf{B})}^\dag = \mathbf{B}^\dag \mathbf{A}^\dag \), implying that \({(\mathbf{A}^n)}^\dag = {(\mathbf{A}^\dag)}^n\), and computing the matrix exponential from its Taylor series.

One can then show that a reversible process (one that satisfies detailed balance and has zero net currents in its steady-state) has a transition matrix that is self-adjoint under \cref{eq:adjoints}, and therefore has real eigenvalues with eigenvectors that are orthogonal under the inner product in \cref{eq:inner}.

\newpage

\section{Primer on large deviation theory for Markov processes}\label{sec:ldt_review}


Here, for the convenience of the general physicist reader,  we outline the derivation of the level 2.5 rate function that is used as a starting point in the main text and \cref{sec:ldt_bnd} of the supplement, at a physical level of rigor.
Much more elaborate proofs can be found in more mathematical works~\cite{Maes2008,Bertini2014,Bertini2015,Barato2015}, which take great care in dealing with subtle issues regarding the existence, uniqueness and convexity of large deviation rate functions.
In our exposition below, we simply present the essential steps and physical intuition, without delving into these  mathematical subtleties.
We hope this provides a straightforward introduction to the large deviation theory of Markov processes for the general physicist.
Below, we first derive the large deviation rate function for empirical densities and fluxes in \cref{sec:fluxLDT}.
Then in \cref{sec:currentLDT} we use the contraction principle to find the rate function for empirical densities and currents.


\subsection{Empirical densities, fluxes, and currents}\label{sec:empirical}

Given a continuous time ergodic Markov process with transition rates \(\MMm_{ij}\) and unique stationary distribution \(\pi_i\), we can imagine observing a particular realization of a trajectory \(\traj \) through a sequence of states \(x_0, x_1, \dots, x_m\) with corresponding transition times \(t_0, t_1, \dots, t_m\).
Each \(x_r \in \{ 1,\dots,n\} \) denotes one of \(n\) possible occupied states of the Markov process.
We can also describe the trajectory in terms of the sequence of states and the  holding time in each state, \(\tau_r = t_{r+1} - t_r\).
This collection of states and holding times \(\trajs \) defines a trajectory, or path \(x(t)\) of the Markov process.
The empirical density for state \(i\) that we would observe over the course of this path is defined (as in the main text) as
\begin{equation} \label{eq:empiricaldensity}
  \pem_i \equiv \frac{1}{T} \int_0^T \dt \, \delta_{x(t) \,i}
    = \frac{\sum_r \delta_{x_r i} \, \tau_r}{T},
\end{equation}
where \(T\) is the duration of the trajectory.
Thus \(\pem_i\) is simply the fraction of time spent in state \(i\) along the trajectory \(\traj \).
Similarly, the empirical flux \(\phiem_{ij} \) is defined as the empirically measured rate at which transitions from state \(i\) to state \(j\) occur over the duration \(T\) of the trajectory \(x(t)\), which we can write as
\begin{equation}\label{eq:empiricalflux}
  \phiem_{ij} = \frac{1}{T} \int_{0}^T \dt \
                    \delta_{x(t^-)i}\, \delta_{x(t^+)j},
    = \frac{\sum_r \delta_{x_r i}\, \delta_{x_{ r+1} j}}{T},
\end{equation}
where \(x(t^-)\) and \(x(t^+)\) are the states before and after a transition.
The empirical \emph{current} from state \(i\) to \(j\) is defined as
\begin{equation}\label{eq:empiricalcurrent}
  \jem_{ij} = \phiem_{ij} - \phiem_{ji},
\end{equation}
\ie the \emph{net} empirical rate of transitions from \(i\) to \(j\) minus \(j\) to \(i\).

All of these observables \(\pem_i\), \(\phiem_{ij}\), and \(\jem_{ij}\) can be measured along a \emph{single} trajectory of the Markov process, and will approach stationary values as the trajectory duration \(T \rightarrow \infty \).
The empirical density will converge to the stationary distribution, \(\pem_i \rightarrow \pi_i\), the empirical flux will converge to the steady state flux, \(\phiem_{ij} \rightarrow \phip_{ij}\), where \(\phip_{ij} = \pi_i \MMm_{ij}\), and the empirical current will converge to the steady state current, \(\jem_{ij} \rightarrow \jp_{ij}\), where \(\jp_{ij} = \pi_i \MMm_{ij} - \pi_j \MMm_{ji}\).

We would now like to find the joint distribution \(\Prob\event{\pem,\phiem}\) over empirical densities \(\pem\) and empirical fluxes \(\phiem\), from which we can find the joint distribution \(\Prob\event{\pem,\jem}\) over empirical densities \(\pem\) and empirical currents \(\jem\).
This will enable us to study the fluctuations and large deviations of these observables around their most likely values of \(\pi \), \(\phip \), and \(\jp \) in the large \(T\) limit.
In this limit, these joint distributions are well characterized by a large deviation rate functions \(I(\pem,\phiem)\) and  \(I(\pem,\jem)\), defined implicitly by the relations
\begin{equation}\label{eq:prob_ldrt}
  \Prob\event{\pem,\phiem} \sim \e^{-T I(\pem,\phiem)},
  \qquad
  \Prob\event{\pem,\jem} \sim \e^{-T I(\pem,\jem)}.
\end{equation}
For an ergodic process \(\MMm \), \(I(\pem,\phiem)\) is expected to be a non-negative convex function of \(\pem\) and  \(\phiem\) that achieves a global minimum value of \(0\) at the unique point \(\pem_i = \pi_i\) and \(\phiem_{ij} = \phip_{ij} = \pi_i \MMm_{ij}\)~\cite{Maes2008}.
Similarly, \(I(\pem,\jem)\) is expected to be a non-negative convex function of \(\pem\) and  \(\jem\) that achieves a global minimum value of \(0\) at the unique point \(\pem_i = \pi_i\) and \(\jem_{ij} = \jp_{ij} = \pi_i \MMm_{ij} - \pi_j \MMm_{jj}\)~\cite{Barato2015,Bertini2014}.
Thus \cref{eq:prob_ldrt} implies that the probability of any large deviation in the empirical density \(\pem_i\), empirical flux \(\phiem_{ij}\), and empirical current \(\jem_{ij}\), from their respective, typical stationary values \(\pi_i\), \(\pi_i \MMm_{ij}\), and \(\pi_i \MMm_{ij} - \pi_j \MMm_{ji}\), is exponentially suppressed in the duration \(T\) of the trajectory.
We present these rate functions in \cref{eq:ldtfluxrate,eq:ldtcurrent} below.
Readers who are simply interested in the form of this function, but not the ideas underlying its derivation, can safely skip the remainder of this appendix.

To derive the rate functions in \cref{eq:prob_ldrt}, we follow the approach of~\cite{Maes2008}.
The essential idea is to use a common tilting method from large deviation theory.
This method involves comparing the probability of a particular path under our given true Markov process with the probability of that same path under a perturbed Markov process in which transition rates are \emph{tilted} to make a particular large deviation more likely.
In particular, we compare the probability of the observed trajectory \(\traj \) under the given Markov process with transition rates \(\MMm_{ij}\), with the probability of the same trajectory under a fictitious process with tilted transition rates \(\widehat{\MMm}_{ij}\).
This fictitious process is the one that would produce the observed trajectory \(x(t)\) as a `typical' realization.
That is, it is a Markov process with stationary distribution \(\widehat{\pi}_i = \pem_i\) and transition rates \(\widehat{\MMm}_{ij} = \phiem_{ij} / \widehat{\pi}_i \).
In essence, empirically densities \(\pem_i\) and fluxes \(\phiem_{ij}\) that might be rarely observed under \(\MMm_{ij}\) are typical under \(\widehat{\MMm}_{ij}\).


\subsection{Trajectory probabilities}\label{sec:traj}

Now, the probability of a particular trajectory \(\traj \) under the original Markov process \(\MMm_{ij}\) is
\begin{align}
\label{eq:probtraj}
  \Prob\events{\traj} = \Prob\event{\trajs}
    &= \Prob\cond{\tau_m \geq T - \tsum_{r=0}^{m-1} \tau_r}{x_m}
       \brk{ \prod_{r=0}^{m-1} \Prob\cond{x_{r+1}}{x_r}
             \Prob\cond{\tau_r}{x_r} } \Prob\event{x_0},
\shortintertext{with}
\label{eq:condP1}
\Prob\cond{x_{r+1}}{x_r} &= \frac{\MMm_{x_r  x_{r+1}}}{\sum_{j} \MMm_{x_r j}}
\shortintertext{and}
\label{eq:condP2}
\Prob\cond{\tau_r}{x_r} &= \lambda_{x_r} \e^{-\lambda_{x_r} \tau_r},
\end{align}
where \(\lambda_{x_r} = \sum_{j\neq x_r} \MMm_{x_r j}\) (see \cref{eq:diags} in \cref{sec:mastereqn}).
\Cref{eq:condP1} is equivalent to \cref{eq:nextstate} which follows from the definition of the transition rates of a continuous time Markov process, and \cref{eq:condP2} is the exponential distribution with parameter \(\lambda_{x_r}\) describing the holding time in each state.

For large \(T\) the boundary effects at \(t=0,T\) will be negligible.
We will neglect those factors in \cref{eq:probtraj} from here on.
The joint probability for the trajectory \(\traj \) is then given by
\begin{equation}\label{eq:ldttraj}
\begin{aligned}
  \Prob\otraj
    &= {\prod_r \Prob\cond{x_{r+1}}{x_r} \Prob\cond{\tau_r}{x_r}} \
    =\ { \prod_{r} \e^{-\lambda_{x_r} \tau_r} \,
                       \MMm_{x_r\, x_{r+1}} }
       \\
    &= \exp\prn{ -\sum_{r=0}^m \lambda_{x_r} \tau_r
                        + \sum_{r=0}^{m-1} \log \MMm_{x_r x_{r+1}}}.
\end{aligned}
\end{equation}

We can now split up the sums over states and transitions indexed by \(r\) into two contributions.
For the first term in the sum in \cref{eq:ldttraj}, we will perform an inner sum over all instances in the trajectory in which the Markov process is in state \(i\), which corresponds to summing over \(r\) such that \(x_r = i\), and then we will perform an outer sum over all of states \(i\) of the Markov process.
Similarly, we will break the second term in \cref{eq:ldttraj} into an inner sum over all of the transitions from state \(i\) to state \(j\) in a trajectory (\ie transitions  in which \(x_r = i, x_{r+1} = j\)), and then an outer sum over all pairs of states \(i\) and \(j\) in the Markov process.
These groupings of sums yield
\begin{equation}
 \Prob\otraj
    = \exp\prn{-\sum_i \sum_{{r : x_r = i}} \lambda_{x_r} \tau_r
    + \sum_{i \neq j} \sum_{{\substack{r : \,x_r=i \\ x_{r+1}=j}}} \log \MMm_{x_r x_{r+1}}}.
\end{equation}
We can now simplify this expression, recognizing that the sum of the occupancy times of state \(i\) during this trajectory is \(\sum_{r : x_r=i} \tau_r =  T \pem_i\), and the total number of transitions from state \(i\) to \(j\) is \(\sum_{\substack{r : \,x_r=i \\ x_{r+1}=j}} 1 = T \phiem_{ij}\):
\begin{equation}\label{eq:ldtregroup}
\begin{aligned}
  \Prob\otraj & = \exp\prn{-\sum_i \lambda_i \sum_{{r : x_r=i}} \tau_r
                + \sum_{i \neq j} \log \MMm_{ij}  \sum_{{\substack{r : \,x_r=i \\ x_{r+1}=j}}} 1} \\
    &= \exp\prn{-\sum_i \lambda_i \, T \pem_i
                + \sum_{i \neq j}T \phiem_{ij} \log \MMm_{ij}}  \\
    &=  \exp\prn{-T \sum_{i \neq j} \brk{\pem_i \MMm_{ij}
                - \phiem_{ij} \log \MMm_{ij}} }.
\end{aligned}
\end{equation}
Thus the probability density assigned to any \emph{individual} trajectory \(\traj \) of duration \(T\) depends on that trajectory \emph{only} through two types of observables: the empirical densities \(\pem_i\) and empirical fluxes \(\phiem_{ij}\).
This dramatic simplification of the distribution over trajectories singles out empirical densities and fluxes as uniquely important order parameters in Markovian non-equilibrium processes.


\subsection{Rate function for densities and fluxes}\label{sec:fluxLDT}

Next, to go from a probability distribution over trajectories \(\traj \) to a joint distribution over empirical densities and fluxes \((\pem,\phiem)\), we would need to integrate over all possible trajectories \(\traj \) that produce any given empirical density and flux pair \((\pem,\phiem)\).
This direct integration would result in adding a difficult to compute entropic contribution to \cref{eq:ldtregroup}.
Instead, we compute the ratio of probabilities for the \emph{same} path under two different processes, \(\MMm \) and the fictitious process with tilted rates \(\widehat{\MMm}\):
\begin{equation}\label{eq:ldtratio}
  \frac{\Prob\otraj}{\widehat{\Prob}\otraj}
     = \exp\prn{-T \sum_{i \neq j} \brk{\pem_i \brk{\MMm_{ij} - \widehat{\MMm}_{ij}}
                + \phiem_{ij} \log\prn{\frac{\widehat{\MMm}_{ij}}{\MMm_{i j}}}}}.
\end{equation}
Noting that this ratio depends on the trajectory \(x(t)\) only though the observables \((\pem,\phiem)\), we can find a computable relation between the distribution \(\Prob\event{\pem,\phiem}\) of these observables under the process \(\MMm \), and the distribution \(\widehat{\Prob}\event{\pem,\phiem}\) of these observables  under the tilted process \(\widehat{\MMm}\):
\begin{equation}\label{eq:ldttrick}
\begin{alignedat}{1}
  \Prob\event{\pem,\phiem}
     &= \int\limits_{\mathrlap{\trajs \to (\pem,\phiem)}} \dtraj\
                \Prob\otraj \
     = \int\limits_{\mathrlap{\trajs \to (\pem,\phiem)}} \dtraj\
                \widehat{\Prob}\otraj \,
                \frac{\Prob\otraj}{\widehat{\Prob}\otraj} \\
     &= \frac{\Prob\otraj}{\widehat{\Prob}\otraj}
                \int\limits_{\mathrlap{\trajs \to (\pem,\phiem)}} \dtraj\
                \widehat{\Prob}\otraj \
     =\ \frac{\Prob\otraj}{\widehat{\Prob}\otraj} \
                \widehat{\Prob}\event{\pem,\phiem},
\end{alignedat}
\end{equation}
where \(\traj \to (\pem,\phiem)\) indicates integration over the set of trajectories that lead to a given empirical density and flux.

For large \(T\), the distributions \(\Prob\event{\pem,\phiem}\) and \(\widehat{\Prob}\event{\pem,\phiem}\) are both characterized by their large deviation rate functions \(I(\pem,\phiem)\) and \(\widehat{I}(\pem,\phiem)\), respectively.
Thus \cref{eq:ldtratio,eq:ldttrick} yield a relation between these two rate functions:
\begin{equation}\label{eq:ldtratediff}
  I(\pem,\phiem) = \widehat{I}(\pem,\phiem) + \sum_{i \neq j}
            \brk{\pem_i \brk{\MMm_{ij} - \widehat{\MMm}_{ij}}
            + \phiem_{ij} \log\prn{\frac{\widehat{\MMm}_{ij}}{\MMm_{ij}}}}.
\end{equation}
So, if we knew the rate function for the fictitious process \(\widehat{\MMm}\), we could find the rate function for \(\MMm \).

Now we note that we have constructed the fictitious process \(\widehat{\MMm}\) \emph{specifically} so that its rate function evaluated at \((\pem,\phiem)\) is exceedingly simple.
Indeed we have chosen the transition rates
\begin{equation*}
  \widehat{\MMm}_{ij} = \frac{\phiem_{ij}}{\pem_i}
\end{equation*}
in order to make the empirically observed densities and fluxes \((\pem,\phiem)\) typical.
Thus, because the observed \((\pem,\phiem)\) are typical under \(\widehat{\MMm}\), we must have \(\widehat{I}(\pem,\phiem) = 0\), so that the probability of these observed values is not exponentially suppressed in \(T\) for large \(T\) under \(\widehat{\MMm}\).
This implies that \cref{eq:ldtratediff} simplifies to yield the sought after expression for the large deviation rate function for joint densities and fluxes under the true process \(\MMm \),
\begin{equation}\label{eq:ldtfluxrate}
  I(\pem,\phiem) = \sum_{i \neq j}
            \brk{{\phipem_{ij} - \phiem_{ij}}
            + \phiem_{ij} \log\prn{\frac{\phiem_{ij}}{\phipem_{ij}}}},
\end{equation}
where we have written \(\phipem_{ij} = \pem_i \MMm_{ij}\).

We note that our notation singles out three types of fluxes:  (a) a completely empirical flux \(\phiem_{ij}\) computed via \cref{eq:empiricalflux} along a single trajectory of duration \(T\) under the process \(\MMm \); (b) a mixed empirical/true flux \(\phipem_{ij} = \pem_i \MMm_{ij}\) computed via the empirical density \(\pem_i\) along a trajectory via \cref{eq:empiricaldensity}, but multiplied by the \emph{true} transition rates \(Q_{ij}\); (c) the true stationary fluxes of the process \(\MMm\) which can be written as \(\phip_{ij} = \pi_i \MMm_{ij}\).
As expected, the large deviation rate function in \cref{eq:ldtfluxrate} is \(0\) if and only if the empirically observed density \(\pem_i\) equals the equilibrium distribution \(\pi_i\) of \(\MMm \) for all states \(i\), and the empirically observed fluxes \(\phiem_{ij}\) equal the stationary fluxes \(\pi_i \MMm_{ij}\) for all pairs of transitions from \(i\) to \(j\).
In this situation, all three fluxes are the same:  \(\phiem_{ij} = \phipem_{ij} = \phip_{ij} = \pi_i \MMm_{ij}\).
However, according to \cref{eq:prob_ldrt} and \cref{eq:ldtfluxrate}, the probability of any discrepancy between any pair of these three fluxes is exponentially suppressed in time \(T\), for large \(T\).


\subsection{Rate function for densities and currents}\label{sec:currentLDT}

The empirical \emph{current} from state \(i\) to \(j\) is defined as \(\jem_{ij} = \phiem_{ij} - \phiem_{ji}\), \ie the net empirical rate of transitions from \(i\) to \(j\) minus \(j\) to \(i\).
To find the joint large deviation rate function for empirical density \(\pem_i\) and current \(\jem_{ij}\), we can apply the contraction principle~\cite{Touchette2009} to the joint rate function for the density and flux in \cref{eq:ldtfluxrate}:
\begin{equation}\label{eq:ldtcontract}
  I(\pem,\jem) = \inf_{\phiem} I(\pem,\phiem)
  \quad\text{ subject to: }
  \phiem_{ij} - \phiem_{ji} = \jem_{ij} \text{ and } \phiem_{ij} \geq 0.
\end{equation}
The infimum can be found by writing \(\phiem_{ij}\) and \(\phiem_{ji}\) in terms of \(\jem_{ij}\) and \(\fem_{ij}\), where we define \(\fem_{ij} \equiv \phiem_{ij} + \phiem_{ji}\).
We can then minimize with respect to \(\fem_{ij}\) subject only to \(\fem_{ij} \geq \abs{\jem_{ij}}\).
The infimum is achieved at \(\fem_{ij} = \sqrt{{(\jem_{ij})}^2 + {(a^p_{ij})}^2}\), where \(a^p_{ij} = 2\sqrt{\phipem_{ij} \phipem_{ji}}\).
If we substitute the flux back into \cref{eq:ldtfluxrate} we find
\begin{equation}\label{eq:ldtcurrent}
\begin{aligned}
  I(\pem,\jem) &=
    \sum_{i<j} \brk{\frac{j^{T}_{ij}}{2}
    \log\prn{\frac{\brk{\fem_{ij} + \jem_{ij}} \brk{\fpem_{ij} - \jpem_{ij}}}
                  {\brk{\fem_{ij} - \jem_{ij}} \brk{\fpem_{ij} + \jpem_{ij}}}}
               - \prn{\fem_{ij} - \fpem_{ij} }}
    \\
     &= \sum_{i<j} \brk{j^{T}_{ij} \prn{\arcsinh{\tfrac{\jem_{ij}}{a^p_{ij}}}
                - \arcsinh{\tfrac{\jpem_{ij}}{a^p_{ij}}}}
               - \prn{\sqrt{j^{T\,2}_{ij} + a^{p\,2}_{ij}}
                    - \sqrt{j^{p\,2}_{ij} + a^{p\,2}_{ij}} }}.
\end{aligned}
\end{equation}
This is the rate function for large deviations of the empirical density and current quoted in the main text (where we have omitted the superscripts $T$ for notational convenience).

\end{appendices}

\vspace{-0.8cm}

\let\oldaddcontentsline\addcontentsline
\renewcommand{\addcontentsline}[3]{}
\let\addcontentsline\oldaddcontentsline

\endgroup

\end{document}